\def\year{2021}\relax
\documentclass[letterpaper]{article} 
\usepackage{aaai21}  
\usepackage{times}  
\usepackage{helvet} 
\usepackage{courier}  
\usepackage[hyphens]{url}  
\usepackage{graphicx} 
\usepackage{natbib}  
\usepackage{caption} 
\usepackage{amsmath}
\usepackage{amsfonts}
\usepackage{amssymb, amsthm}
\usepackage{subcaption}
\usepackage{mathtools} 

\usepackage{xcolor}
\usepackage{color, colortbl}
\usepackage{soul}
\usepackage{stmaryrd}  
\usepackage{xcolor}  
\usepackage[switch]{lineno}  
\usepackage{tikz}
\usepackage{qtree}
\tikzstyle{input} = [rectangle, minimum width=2cm, minimum height=1cm,text centered, draw=black, fill=green!30]
\tikzstyle{result} = [rectangle, minimum width=2cm, minimum height=1cm,text centered, draw=black, fill=red!30]
\tikzstyle{stage} = [rectangle, rounded corners, minimum width=3cm, minimum height=1cm,text centered, draw=black, fill=blue!30]
\tikzstyle{decision} = [diamond, minimum width=2cm, minimum height=2cm, text centered, draw=black, fill=orange!30]
\input{algorithm.tex} 

\title{Learning of Structurally Unambiguous Probabilistic Grammars}
\author{Dolav Nitay\thanks{This research  was partially funded by BSF Grant  \#2016239, ISF Grant \#939/18  \&  Frankel Center for Computer Science, BGU.}
, Dana Fisman, Michal Ziv-Ukelson \\}
\affiliations {
dolavn@post.bgu.ac.il\quad dana@cs.bgu.ac.il\quad michaluz@bgu.ac.il\\[2.5mm]
Ben Gurion University
}
\begin{document}
\newcommand{\commentout}[1]{}

\newcommand{\dana}[1]{\colorbox{yellow}{......}\footnote{\colorbox{yellow}{\textsc{df: [}}#1{\colorbox{yellow}{]}}}}
\newcommand{\dolav}[1]{\colorbox{red}{......}\footnote{\colorbox{red}{\textsc{dn: [}}#1{\colorbox{red}{]}}}}
\newcommand{\michal}[1]{\colorbox{green}{......}\footnote{\colorbox{green}{\textsc{mz: [}}#1{\colorbox{green}{]}}}}

\newcommand{\dolavlong}[2]{{\colorbox{red}{Dolav: #1 }\\ {\colorbox{red}{ [\quad }\\ {#2} \\{\colorbox{red}{ \quad] } }\\}}}
\newcommand{\danalong}[2]{{\colorbox{yellow}{Dana: #1 }\\ {\colorbox{yellow}{ [\quad }\\ {#2} \\{\colorbox{yellow}{ \quad] } }\\}}}
\newcommand{\michallong}[2]{{\colorbox{green}{Michal: #1 }\\ {\colorbox{green}{ [\quad }\\ {#2} \\{\colorbox{green}{ \quad] } }\\}}}

\newcommand{\parciteauthor}[1]{(\citeauthor{#1})}

\newcommand{\reals}{\mathbb{R}}

\newcommand{\biword}[2]{\ensuremath{\left(\begin{smallmatrix} #1 \\ #2 \end{smallmatrix}\right)}}

\newcommand{\grmrwgt}[2]{\ensuremath{\mathcal{W}_{#1}(#2)}}

\newcommand{\query}[1]{\textsc{#1}}
\newcommand{\mq}{\query{mq}}
\newcommand{\smq}{\query{smq}}
\newcommand{\eq}{\query{eq}}
\newcommand{\seq}{\query{seq}}

\newcommand{\complexityclass}[1]{\textbf{#1}}
\newcommand{\RP}{\complexityclass{RP}}
\newcommand{\NP}{\complexityclass{NP}}

\newcommand{\Class}[1]{\mathbb{W}_{#1}}
\newcommand{\ClassT}[1]{\mathbb{T}_{#1}}

\newcommand{\alg}[1]{\ensuremath{\textbf{#1}}}
\newcommand{\staralg}[1]{\ensuremath{{\textbf{#1}^*}}}
\newcommand{\lstar}{\staralg{L}}
\newcommand{\gstar}{\staralg{G}}
\newcommand{\mstar}{\staralg{M}}
\newcommand{\pstar}{\staralg{P}}
\newcommand{\cstar}{\staralg{C}}

\newcommand{\deriv}{\ensuremath{\mathsf{T}}}
\newcommand{\struct}{\ensuremath{\mathsf{S}}}
\newcommand{\skels}{\ensuremath{\mathsf{S}}}
\newcommand{\skel}{\ensuremath{\mathsf{S}}}
\newcommand{\skelstotrees}{\ensuremath\mathsf{T}}

\newcommand{\aut}[1]{\mathcal{#1}}
\newcommand{\grmr}[1]{\ensuremath{\mathcal{#1}}}
\newcommand{\treesrs}[1]{\ensuremath{\mathcal{#1}}}

\newcommand{\treesEquiv}[1]{\equiv_{#1}}
\newcommand{\treesColin}[2]{\ltimes^{#1}_{#2}}
\newcommand{\classrepr}[2]{[#1]}
\newcommand{\classcoeff}[2]{\alpha[#1]}
\newcommand{\classind}[2]{\iota[#1]}

\newcommand{\tuple}[1]{\ensuremath\langle #1 \rangle}

\newcommand{\derives}{\xrightarrow{}}
\newcommand{\transderives}[1]{\ensuremath{\Rightarrow_{#1}}}
\newcommand{\Vars}{\ensuremath{\mathcal{V}}}

\newcommand{\norm}[1]{\ensuremath{\|{#1}\|}}
\newcommand{\reduce}[1]{\ensuremath{\llbracket{#1}\rrbracket}}
\newcommand{\restrict}[2]{#1\mid_{#2}}
\newcommand{\restrictrowscols}[3]{{#1}[{#2},{#3}]}
\newcommand{\restrictrows}[2]{{#1}[{#2},*]}
\newcommand{\restrictcols }[2]{{#1}[*,{#2}]}
\newcommand{\restrictrow}[2]{{#1}|_{#2}}

\newcommand{\Prob}{\mathbb{P}}

\newtheorem{theorem}{Theorem}[section]
\newtheorem{proposition}[theorem]{Proposition}
\newtheorem{corollary}[theorem]{Corrolary} 
\newtheorem{lemma}[theorem]{Lemma} 
\newtheorem{claim}[theorem]{Claim} 
\newtheorem{definition}[theorem]{Definition} 
\newtheorem{notation}[theorem]{Notations} 

\newcommand{\claimref}[1]{\textbf{#1}.}

\newcommand{\F}{\mathcal{F}}
\newcommand{\V}{\mathbb{V}}
\newcommand{\K}{\mathbb{K}}
\newcommand{\R}{\mathbb{R}}

\newcommand{\hm}[1]{\ensuremath{{#1}}}

\newcommand{\rank}{\textsl{rank}}
\newcommand{\crank}{\ensuremath{\textsl{c-rank}_{+}}}
\newcommand{\rrank}{\ensuremath{\textsl{r-rank}_{+}}}
\newcommand{\prank}{\ensuremath{\textsl{rank}_{+}}}
\newcommand{\trees}{\textsl{Trees}}
\newcommand{\ts}{\textsl{TS}}
\newcommand{\pos}{\textsl{span}_{+}}
\newcommand{\posspan}[1]{\pos(#1)}

\newcommand{\proc}[1]{\textsl{#1}}
\newcommand{\PositiveBase}{\proc{PositiveBase}}
\newcommand{\ExtractCMTA}{\proc{ExtractCMTA}}
\newcommand{\LearnCMTA}{\proc{LearnCMTA}}
\newcommand{\Close}{\proc{Close}}
\newcommand{\Consistent}{\proc{Consistent}}
\newcommand{\Complete}{\proc{Complete}}
\newcommand{\CloseAll}{\proc{CloseAll}}
\newcommand{\Rebase}{\proc{Rebase}}

\newcommand{\sema}[1]{{\llbracket}#1{\rrbracket}}%
\newcommand{\semagrmr}[1]{#1}%

\newcommand{\rmi}{\textsc{i}}
\newcommand{\rmii}{\textsc{ii}}

\newcommand{\context}{\diamond}
\newcommand{\suff}{\textsl{Suff}}
\newcommand{\pref}{\textsl{Pref}}

\newcommand{\contextConcat}[2]{#1\llbracket#2\rrbracket}
\newcommand{\yield}{\textsl{yield}}
\newcommand{\opt}{\textsl{OPT}}

\newcommand{\AlgClose}{2}
\newcommand{\AlgComplete}{4}
\newcommand{\AlgLearnCmta}{1}
\newcommand{\AlgConsistent}{3}
\newcommand{\AlgExtractCmta}{5}

\newcommand{\AppExample}{A}
\newcommand{\AppProofs}{B}
\newcommand{\AppBio}{C} 
\maketitle
{
\begin{abstract}
    	The problem of identifying a probabilistic context free grammar has two aspects: the first is determining the grammar's topology (the rules of the grammar) and the second is estimating probabilistic weights for each rule. Given the hardness results for learning context-free grammars in general, and probabilistic grammars in particular, most of the literature has concentrated on the second problem. In this work we address the first problem.  We restrict attention to structurally unambiguous weighted context-free grammars (SUWCFG) and provide a query learning algorithm for strucuturally unambiguous probabilistic context-free grammars  (SUPCFG). We show that SUWCFG can be represented using co-linear multiplicity tree automata (CMTA), and provide a polynomial learning algorithm that learns CMTAs.  We show that the learned CMTA can be converted into a probabilistic grammar, thus providing  a complete algorithm for learning  a strucutrally unambiguous probabilistic context free grammar (both the grammar topology and the probabilistic weights) using structured membership queries and structured equivalence queries. 
    	We demonstrate the usefulness of our algorithm in learning PCFGs over genomic data.
\end{abstract}
}


{

\section{Introduction}\label{sec:intro}

Probabilistic context free grammars (PCFGs) constitute a computational model suitable for probabilistic systems which observe non-regular (yet context-free) behavior. They are vastly used in computational linguistics~\parciteauthor{Chomsky56}, natural language processing~\parciteauthor{church-1988-stochastic} and biological modeling, for instance, in probabilistic modeling of RNA structures~\parciteauthor{Grate95}.  Methods for learning PCFGs from experimental data have been thought for over half a century.
Unfortunately, there are various hardness results regarding learning context-free grammars in general and probabilistic grammars in particular. 
It follows from~\parciteauthor{Gold78}  that context-free grammars (CFGs) cannot be identified in the limit from positive examples, and from~\cite{Angluin90}  that CFGs cannot be identified in polynomial time using equivalence queries only. Both results are not surprising for those familiar with learning regular languages, as they hold for the class of regular languages as well. However, while regular languages can be learned using both membership queries and equivalence queries~\cite{Angluin87}, it was shown   that learning CFGs using both membership queries and equivalence queries  
is computationally as hard as key cryptographic problems for which there is currently no known polynomial-time algorithm~\parciteauthor{AngluinK95}. {See more on the difficulties of learning context-free grammars in~\cite[Chapter 15]{delaHiguera}.}
Hardness results for the probabilistic setting have also been established. \parciteauthor{AbeW92} have shown a computational hardness result for the inference of probabilistic automata, in particular, that an exponential blowup with respect to the alphabet size is inevitable unless $\RP = \NP$. 

The problem of identifying a probabilistic grammar from examples has two aspects: the first is determining the rules of the grammar up to variable renaming and the second is estimating probabilistic weights for each rule. Given the hardness results mentioned above, most of the literature has concentrated on the second problem. Two dominant approaches for solving the second problem are the
forward-backward algorithm for HMMs~\parciteauthor{Rabiner89} and the inside-outside algorithm for PCFGs~\parciteauthor{Baker79,LariY90}.

In this work we address the first problem. Due to the hardness results regarding learning probabilistic grammars using membership and equivalence queries (\mq\ and \eq) we use structured membership queries and structured equivalence queries (\smq\ and \seq), as was done by~\cite{Sakakibara88} for learning context-free grammars. \emph{Structured strings}, proposed by~\parciteauthor{LevyJ78}, are strings over the given alphabet that includes parentheses that indicate the structure of a possible derivation tree for the string. One can equivalently think about a structured string as a derivation tree in which all nodes but the leaves are marked with $?$, namely an \emph{unlabeled derivation tree}. 

It is known that the set of derivation trees of a given CFG constitutes a \emph{regular tree-language}, where a regular tree-language is a tree-language that can be recognized by a \emph{tree automaton}. 
\cite{Sakakibara88} has generalized Angluin's \lstar\ algorithm (for learning regular languages using \mq\ and \eq) to learning a tree automaton, and provided a polynomial learning algorithm for CFGs using \smq\ and \seq.
Let $\deriv(\grmr{G})$ denote the set of derivation trees of a CFG $\grmr{G}$, and $\struct(\deriv(\grmr{G}))$ the set of unlabeled derivation trees (namely the structured strings of $\grmr{G}$). 
While a membership query (\mq) asks whether a given string $w$ is in the unknown grammar $\grmr{G}$, 
a structured membership query (\smq) asks whether a structured string $s$ is in   $\struct(\deriv(\grmr{G}))$ and a structured equivalence query (\seq) answers whether the queried CFG $\grmr{G}'$ is structurally equivalent to the unknown grammar $\grmr{G}$, and accompanies a negative answer with a structured string $s'$ in the symmetric difference of  $\struct(\deriv(\grmr{G}'))$ and $\struct(\deriv(\grmr{G}))$. 

In our setting, since we are interested in learning probabilistic grammars, an \smq\ on a structured string $s$ is answered by a weight $p\in[0,1]$ standing for the probability for $\grmr{G}$ to generate $s$, and a negative answer to an \seq\ is accompanied by a structured string $s$ such that $\grmr{G}$ and $\grmr{G}'$ generate $s$ with different probabilities (up to a predefined error margin) along with  the probability $p$ with which the unknown grammar $\grmr{G}$ generates $s$.

\cite{Sakakibara88} works with tree automata to model the derivation trees of the unknown grammars.
In our case the automaton needs to associate a weight with every tree (representing a structured string). We choose to work with the model of \emph{multiplicity tree automata}. A multiplicity tree automaton (MTA) associates with every tree a value from a given field $\K$. An algorithm for learning multiplicity tree automata, to which we refer as \mstar, was developed in~\parciteauthor{habrard2006learning,drewes2007query}.\footnote{Following a learning algorithm developed for multiplicity word automata~\parciteauthor{BergadanoV96}.} 

A probabilistic grammar is a special case of a weighted grammar and~\parciteauthor{abney1999relating,smith2007weighted} have shown that  convergent weighted CFGs (WCFG) where all weights are non-negative and probabilistic CFGs (PCFGs) are equally expressive.\footnote{The definition of \emph{convergent} is deferred to the preliminaries.} We thus might expect to be able to use the learning algorithm \mstar\ to learn an MTA corresponding to a WCFG, and apply this conversion  to the result, in order to obtain the desired PCFG. However, as we show in  Proposition~\ref{prop:needPMTAs}, there are probabilistic languages for which applying the \mstar\ algorithm results in an MTA with negative weights. Trying to adjust the algorithm to learn a positive basis may encounter the issue that for some PCFGs,  no finite subset of the infinite Hankel Matrix spans the entire space of the function, as we show in Proposition~\ref{prop:pcfg-no-finite-rank}.\footnote{The definition of the Hankel Matrix and its role in learning algorithms appears in the sequel.}  To overcome these issues we restrict attention to structurally unambiguous grammars (SUCFG, see section \ref{sec:sucfg}), which as we show, can be modeled using co-linear multiplicity automata (defined next).

We develop a polynomial learning algorithm, which we term \cstar, that learns a restriction of MTA, which we term \emph{co-linear multiplicity tree automata} (CMTA). We then show that a CMTA for a probabilistic language can be converted into a PCFG, thus yielding a complete algorithm for learning SUPCFGs using \smq s and \seq s as desired.

As a proof-of-concept, in Section \ref{sec:demonstration} we exemplify our algorithm by applying it to a small data-set of genomic data. 

Due to lack of space all proofs are deferred to appendix (App.~\AppProofs). The appendix also contains (i) a complete running example (App.~\AppExample) and (ii)
supplementary material for the demonstration section (App.~\AppBio).

\section{Preliminaries}

This section provides the  definitions required for  \emph{probabilistic grammars} -- the object we design a learning algorithm for,
and \emph{multiplicity tree automata}, the object we use in the learning algorithm.
\subsection{Probabilistic Grammars}\label{subsec:pcfgs}

Probabilistic grammars are a special case of context free grammars 
where each production rule has a weight in the range $[0,1]$ and for each non-terminal, the sum of weights of its productions is one. 
%
	A \emph{context free grammar} (CFG) is a quadruple $\grmr{G}=\langle \Vars,\Sigma,R,S\rangle$, where
	$\Vars$ is a finite non-empty set of symbols called  \emph{variables} or \emph{non-terminals},
	$\Sigma$ is a finite non-empty set of symbols called the \emph{alphabet} or the \emph{terminals},
	$R\subseteq \Vars\times (\Vars\cup\Sigma)^{*}$ is a relation between variables and strings over $\Vars\cup\Sigma$, called the  \emph{production rules}, and $S\in \Vars$ is a special variable called the \emph{start variable}. 
	We assume the reader is familiar with the standard definition of CFGs and of derivation trees.
%
We say that $S\transderives{} w$ for a string $w\in\Sigma^*$ if there exists a derivation tree $t$ such that all leaves are in $\Sigma$ and when concatenated from left to right they form $w$. That is, $w$ is the \emph{yield} of the tree $t$. In this case we also use the notation $S\transderives{t} w$.
A CFG $\grmr{G}$ defines a set of words over $\Sigma$, the \emph{language generated by} $\grmr{G}$, which  is the set of words $w\in\Sigma^*$ such that $S\transderives{}  w$, and is denoted $\sema{\grmr{G}}$. For simplicity, we assume the grammar does not derive the empty word.

\vspace{-2mm}
\paragraph{Weighted grammars}
A \emph{weighted grammar}  (WCFG) is a pair $\tuple{\grmr{G},\theta}$ where $\grmr{G}=\langle \Vars,\Sigma,R,S\rangle$ is a CFG and $\theta:R\rightarrow \R$ is a function mapping each production rule to a weight in $\R$.
A WCFG $\grmr{W}=\tuple{\grmr{G},\theta}$ defines a function from words over $\Sigma$ to weights in $\R$. 
The WCFG associates with a derivation tree $t$ its weight, which is defined as 
$\grmr{W}(t)=\prod_{(V\derives \alpha)\in R } \theta(V\derives\alpha)^{\sharp_t(V\derives\alpha)}$
where $\sharp_t(V\derives\alpha)$ is the number of occurrences of the production $V\derives\alpha$ in the derivation tree $t$.
We abuse notation and treat $\semagrmr{\grmr{W}}$ also as a function from $\Sigma^*$ to $\R$  defined as 
$\grmr{W}(w)=\sum_{S\transderives{t}w}\grmr{W}(t)$.
That is, the weight of $w$ is the sum of weights of the derivation trees yielding $w$, and if $w\notin\sema{\aut{G}}$ then $\semagrmr{\grmr{W}}(w)=0$.
If the sum of all derivation trees in $\sema{\grmr{G}}$, namely $\sum_{w\in\sema{\grmr{G}}}\grmr{W}(w)$, is finite we say that $\aut{W}$ is \emph{convergent}. 

\vspace{-2mm}
\paragraph{Probabilistic grammars}
A \emph{probabilistic grammar} (PCFG) is a WCFG 
$\grmr{P}=\tuple{\grmr{G},\theta}$ where $\grmr{G}=\langle \Vars,\Sigma,R,S\rangle$ is a CFG and $\theta:R\rightarrow [0,1]$ is a function mapping each production rule of $\grmr{G}$ to a weight in the range $[0,1]$ that satisfies 
$1=\sum_{(V \derives \alpha_i) \in R} \theta(V\derives \alpha_i) $ for every $V\in \Vars$.\footnote{Probabilistic grammars are sometimes called \emph{stochastic grammars (SCFGs)}.}
One can see that if $\grmr{P}$ is a PCFG then  the sum of all derivations equals $1$, thus $\grmr{P}$  is convergent.

\subsection{Word/Tree Series and Multiplicity Automata}\label{sec:mta}
While words are defined as sequences over a given alphabet, trees are defined using a \emph{ranked alphabet},
an alphabet $\Sigma=\{\Sigma_0,\Sigma_1,\ldots,\Sigma_p\}$ which is a  tuple of alphabets $\Sigma_k$ where $\Sigma_0$ is non-empty.
Let $\trees(\Sigma)$ be the set of trees over $\Sigma$, where a node labeled $\sigma\in\Sigma_k$ for $0\leq k \leq p$ has exactly $k$ children. While a \emph{word language} is a function mapping all possible words (elements of $\Sigma^*$) to $\{0,1\}$, a \emph{tree language} is a function from all possible trees (elements of $\trees(\Sigma)$) to $\{0,1\}$. We are interested in assigning each word or tree a non-Boolean value, usually a weight $p\in[0,1]$. Let $\K$ be a field. We are interested in functions mapping words or trees to values in $\K$. A function from $\Sigma^*$ to $\K$ is  called a \emph{word series}, and a function from $\trees(\Sigma)$ to $\K$ is referred to as a \emph{tree series}.

\emph{Word automata} are machines that recognize word languages, i.e. they define a function from $\Sigma^*$ to $\{0,1\}$.
\emph{Tree automata} are machines that recognize tree languages, i.e. they define a function from $\trees(\Sigma)$ to $\{0,1\}$.
\emph{Multiplicity word automata} (MA) are machines to implement word series, i.e. they define a function from $\Sigma^*$ to $\K$.
\emph{Multiplicity tree automata} (MTA) are machines to implement tree series, i.e. they define a function from $\trees(\Sigma)$ to $\K$.
Multiplicity automata can be thought of as an algebraic extension of automata, in which reading an input letter is implemented by matrix multiplication.
In a multiplicity word automaton with dimension $m$ over alphabet $\Sigma$, for each $\sigma\in\Sigma$ there is an $m$ by $m$ matrix, $\mu_\sigma$, whose entries are values in $\K$ where intuitively the value of entry $\mu_\sigma(i,j)$ is the weight of the passage from state $i$ to state $j$. The definition of multiplicity tree automata is a bit more involved; it makes use of multilinear functions as defined next. 

\vspace{-2mm}
\paragraph{Multilinear functions}\label{par:multilinear functions}
Let $\V=\K^d$ be the $d$ dimensional vector space over $\K$. Let $\eta : \V^k \rightarrow \V$ be a $k$-linear function.
We can represent $\eta$ by a $d$ by $d^k$ matrix over $\K$. For instance, if $\eta : \V^3 \rightarrow \V$ and $d=2$ (i.e. $\V = \K^2$) then $\eta$ can be represented by the $2\times 2^3$ matrix $M_\eta$ provided in Fig~\ref{fig:mult-aut-matrices} where $c^{i}_{j_1 j_2 j_3} \in \K$ for $i,j_1,j_2,j_3\in\{1,2\}$. 
Then $\eta$, a function taking $k$ parameters in $\V=\K^d$, can be computed by multiplying the matrix $M_\eta$ with a vector for the parameters for $\eta$. Continuing this example, given the parameters $\textbf{x} = (x_1 \ \ x_2)$, $\textbf{y} = (y_1 \  \ y_2)$, $\textbf{z} = (z_1\ \  z_2)$ the value
$\eta(\textbf{x},\textbf{y},\textbf{z})$ can be calculated using the multiplication $M_\eta P_{xyz}$ where the vector $P_{xyz}$  of size $2^3$ is provided in Fig~\ref{fig:mult-aut-matrices}.
\begin{figure}
	\scalebox{.7}{\small{
			\begin{tabular}{l@{\quad}l}
				$M_\eta = \begin{pmatrix}
				c^1_{{111}} & c^1_{{112}}& c^1_{{121}} & c^1_{{122}} & c^1_{{211}} & c^1_{{212}}& c^1_{{221}} & c^1_{{222}} \\[1mm]
				c^2_{{111}} & c^2_{{112}}& c^2_{{121}} & c^2_{{122}} & c^2_{{211}} & c^2_{{212}}& c^2_{{221}} & c^2_{{222}} 
				\end{pmatrix}$
				&
				$P_{xyz} =  \begin{pmatrix}
				x_1 y_1 z_1 \\
				x_1 y_1 z_2 \\
				x_1 y_2 z_1 \\
				\ldots \\
				x_2 y_2 z_2 \\
				\end{pmatrix}$
			\end{tabular}
	}}
	\commentout{
	\scalebox{.8}{\small{
			\begin{tabular}{l@{\quad}l}
				$M_\eta = \begin{pmatrix}
				c^1_{{111}} & c^1_{{112}}& c^1_{{121}} & c^1_{{122}} & c^1_{{211}} & c^1_{{212}}& c^1_{{221}} & c^1_{{222}} \\[1mm]
				c^2_{{111}} & c^2_{{112}}& c^2_{{121}} & c^2_{{122}} & c^2_{{211}} & c^2_{{212}}& c^2_{{221}} & c^2_{{222}} 
				\end{pmatrix}$
				&
				$P_{xyz} =  \begin{pmatrix}
				x_1 y_1 z_1 \\
				x_1 y_1 z_2 \\
				x_1 y_2 z_1 \\
				\ldots \\
				x_2 y_2 z_2 \\
				\end{pmatrix}$
			\end{tabular}
	}}}
	\caption{A matrix $M_\eta$ for a multi-linear fuction $\eta$ and a vector $P_{xyz}$ for the respective $3$ parameters.
	}\label{fig:mult-aut-matrices}
\end{figure} 
In general, if $\eta : \V^k \rightarrow \V$ is such that $\eta(\textbf{x}_1,\textbf{x}_2,\ldots,\textbf{x}_k) = \textbf{y}$ and $M_\eta$, the matrix representation of $\eta$,  is defined using the constants $c^{i}_{j_1 j_2 \ldots j_k}$ then 

\[\textbf{y}[i] = \hspace{-1em} \sum_{\left\{(j_1,j_2,\ldots,j_k)\in \{1,2,\ldots,d\}^k\right\}} \hspace{-1em} c^i_{j_1j_2\ldots j_k} \, \textbf{x}_1[j_1]\, \textbf{x}_2[j_2]\, \cdots \,\textbf{x}_k[j_k]\phantom{------}\qquad\qquad\]
\begin{figure} 
	\scalebox{.8}{
	    \hspace{10mm}\includegraphics[scale=0.8,page=8, clip, trim=2cm 22.5cm 9cm 1.8cm]{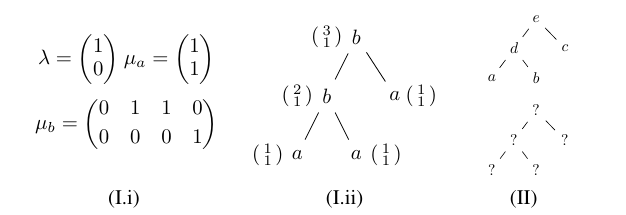}
		\commentout{
			\begin{tabular}{cc}
				$\lambda=\begin{pmatrix}1\\0
				\end{pmatrix}$ \ \ \ 
				$\mu_{a}= \begin{pmatrix}1\\1
				\end{pmatrix}$\\ \quad\vspace{0mm} \\
				$\mu_{b} = \begin{pmatrix}
				0 & 1 & 1 & 0 \\[1mm]
				0 & 0 & 0 & 1 
				\end{pmatrix}$
				
				\\ \quad\vspace{3mm} \\
				\begin{tikzpicture}
				\tikzstyle{myarrow}=[line width=.5mm,draw=#1,-triangle 45,postaction={draw, line width=4mm, shorten >=5mm, -}]
				\node[shape=circle, draw=none] (a1) at (2, 1) {$a$};
				\node[] (a1v) at (3.5, 1) {${\biword{1}{1}}$};
				\node[] (a2v) at (1.5, 1) {${\biword{1}{1}}$};
				\node[shape=circle, draw=none]  (a2) at (3, 1) {$a$};
				\node[shape=circle, draw=none] (a3) at (3.65, 2.0) {$a$};
				\node[] (a3v) at (4.10, 2.0) {${\biword{1}{1}}$};
				\node[shape=circle, draw=none]  (b1) at (2.5, 2.0) {$b$};
				\node[] (a3v) at (2.0, 2.0) {${\biword{2}{1}}$};
				\node[shape=circle, draw=none]  (b2) at (3, 3) {$b$};
				\node[] (b2v) at (2.5, 3) {${\biword{3}{1}}$};
				\draw (b2) -- (a3);
				\draw (b2) -- (b1);
				\draw (b1) -- (a1);
				\draw (b1) -- (a2);
				\end{tikzpicture}
	\end{tabular}}}
	\caption{(I.i) An MTA $\aut{M}=((\Sigma_0,\Sigma_2),\R,2,\mu,\lambda)$
		where    ${\Sigma_0=\{a\}}$ and ${\Sigma_2=\{b\}}$  implementing a tree series that returns the number of leaves in the tree. 
		(I.ii) a tree where a node $t$ is annotated by $\mu(t)$. Since $\mu(t_\epsilon)=\biword{3}{1}$, 
		where $t_\epsilon$ is the root, the value of the entire tree is  $\lambda\cdot\biword{3}{1}=3$.
	 (II) a derivation tree and its corresponding skeletal tree.}\label{fig:MTA}\label{fig:structured-string}
\end{figure}

\paragraph{Multiplicity tree automata}
A \emph{multiplicity tree automaton} (MTA) is a tuple $\aut{M}=(\Sigma,\K,d,\mu,\lambda)$ where $\Sigma=\{\Sigma_0,\Sigma_1,\ldots,\Sigma_p\}$ is the given ranked alphabet, $\K$ is the respective field, $d$ is a non-negative integer called the  automaton \emph{dimension}, $\mu$ and $\lambda$ are  the transition and output function, respectively, whose types are defined next.
Let $\V=\K^d$. Then  $\lambda$ is an element of $\V$, namely a $d$-vector over $\K$.  Intuitively, $\lambda$ corresponds to the final  values of the ``states'' of $\aut{M}$. The transition function $\mu$ maps each element $\sigma$  of $\Sigma$ to a dedicated transition function $\mu_\sigma$ such that given ${\sigma \in \Sigma_k}$ for ${0 \leq k \leq p}$ then $\mu_\sigma$ is a $k$-linear function from $\V^k$ to $\V$. 
The transition function $\mu$ induces a function from $\trees(\Sigma)$ to $\V$, defined as follows.
If $t=\sigma$ for some $\sigma\in \Sigma_0$, namely $t$ is a tree with one node which is a leaf, then $\mu(t)=\mu_\sigma$ (note that $\mu_\sigma$ is a vector in $\K^d$ when $\sigma\in\Sigma_0$). 
If $t=\sigma(t_1,\ldots,t_k)$, namely $t$ is a tree with root $\sigma\in \Sigma_k$ and children $t_1,\ldots,t_k$ then $\mu(t)=\mu_\sigma(\mu(t_1),\ldots,\mu(t_k))$. 
The automaton $\aut{M}$ induces a total function from $\trees(\Sigma)$ to $\K$ defined as follows $\aut{M}(t)=\lambda\cdot \mu(t)$.
Fig.~\ref{fig:MTA}(I.i) provides an example of an MTA and the value for a computed tree Fig.~\ref{fig:MTA}(I.ii).

\paragraph{Contexts}
In the course of the algorithm we need a way to compose trees, more accurately we compose trees with \emph{contexts} as defined next.
Let $\Sigma=\{\Sigma_0,\Sigma_1,\ldots,\Sigma_p\}$ be a ranked alphabet. Let $\context$ be a symbol not in $\Sigma$. 
We use $\trees_\context(\Sigma)$ to denote all non-empty trees over $\Sigma' = \{ \Sigma_0 \cup \{\context\}, \Sigma_1,\ldots, \Sigma_p\}$  in which $\context$ appears exactly once.
We refer to an element of $\trees_\context(\Sigma)$ as a \emph{context}. Note that  at most one child of any node in a context $c$ is a context; the other ones are pure trees (i.e. elements of $\trees(\Sigma))$.
Given a tree ${t\in \trees(\Sigma)}$ and context $c\in \trees_\context(\Sigma)$ we use $\contextConcat{c}{t}$ for the tree $t'\in\trees(\Sigma)$ obtained from $c$ by replacing $\context$ with $t$.

\paragraph{Structured tree languages/series}
Recall that our motivation is to learn a word (string) series rather than a tree series, and due to hardness results on learning CFGs and PCFGs we resort to using \emph{structured strings} which are strings with parentheses exposing the structure of a derivation tree for the corresponding trees. 
These are defined formally as follows (and depicted in Fig.~\ref{fig:structured-string} (II)).
A \textit{skeletal alphabet} is a ranked alphabet in which we use a special symbol $?\notin \Sigma_0$ and for every $0<k\leq p$ the set $\Sigma_{k}$ consists only of the symbol $?$. Let $t\in\trees(\Sigma)$, the skeletal description of $t$, denoted by $\skel(t)$, is a tree with the same topology as $t$, in which the symbol in all internal nodes is $?$, and the symbols in all leaves are the same as in $t$. Let $T$ be a set of trees. The corresponding skeletal set, denoted $\skels(T)$ is $\{\skels(t)~|~t\in T\}$.
Going from the other direction, given a skeletal tree $s$ we use $\skelstotrees(s)$ for the set $\{t\in T~|~\skel(t)= s\}$.

A tree language over a skeletal alphabet is called a \textit{skeletal tree language}. And a mapping from skeletal trees to $\K$ is called a \emph{skeletal tree series}. 
Let $\treesrs{T}$  denote a tree series mapping trees in $\trees(\Sigma)$ to $\K$. 
By abuse of notations, given a skeletal tree, we use $\treesrs{T}(s)$ for the sum of values $\treesrs{T}(t)$ for every tree $t$ of which $s=\skel(t)$.
That is,
$\treesrs{T}(s)=\sum_{t\in\skelstotrees(s)}\treesrs{T}(t)$.
Thus, given a tree series $\treesrs{T}$ (possibly generated by a WCFG or an MTA) we can treat $\treesrs{T}$ as a skeletal tree series.

\section{From Positive MTAs to PCFGs}\label{sec:CMTA2PCFG}
 
Our learning algorithm for probabilistic grammars builds on the relation between WCFGs with positive weights and PCFGs~\parciteauthor{abney1999relating,smith2007weighted}. In particular, we first establish that a \emph{positive multiplicity tree automaton} (PMTA), which is a multiplicity tree automaton (MTA) where all weights are positive, can be transformed into an equivalent  WCFG $\grmr{W}$.
That is, we  show that  a given PMTA $\aut{A}$  over a skeletal alphabet can be converted into a WCFG $\grmr{W}$ such that for every structured string $s$ we have that $\aut{A}(s)=\aut{W}(s)$. 
If the PMTA defines a convergent tree series (namely the sum of weights of all trees is finite) then so will the constructed WCFG. 
Therefore, given that the WCFG describes a probability distribution, we can apply the transformation of WCFG to a PCFG~\parciteauthor{abney1999relating,smith2007weighted} to yield a PCFG $\grmr{P}$ such that $\grmr{W}(s)=\grmr{P}(s)$, obtaining the desired PCFG for the unknown tree series.

\paragraph{Transforming a PMTA into WCFG}
Let $\aut{A}=(\Sigma,\mathbb{R}_+,d,\mu,\lambda)$ be a PMTA over the skeletal alphabet $\Sigma=\{\Sigma_0,\Sigma_1,\ldots,\Sigma_p\}$. We define a WCFG $\grmr{W}_\aut{A}=(\grmr{G}_\aut{A},\theta)$ for $\grmr{G}_{\aut{A}}=(\Vars,\Sigma_{0},R,S)$ as provided in Fig.~\ref{fig:eqs-pmta-to-pcfg} where $c^i_{i_{1},i_{2},...,i_{p}}$ is the respective coefficient in the matrix corresponding to $\mu_?$ for $?\in\Sigma_p$.
\begin{figure}[t]
	\centering
	\scalebox{.85}{	
		{\framebox[1\width]
				{\small{
					$\begin{array}{l@{ }l}
					\Vars= \{S\}\cup\{V_{i}~|~1\leq i\leq d\} &    \\
					R =  \{  S\rightarrow V_{i} ~|~1\leq i\leq d\}\ \cup &   \phantom{..} \theta( S\rightarrow V_{i}) = \lambda[i] \\
					\phantom{R = }\{  V_{i}\rightarrow \sigma  ~|~1\leq i\leq d,\ \sigma\in\Sigma_0 \}\  \cup  &  \phantom{..} \theta( V_{i}\rightarrow \sigma ) = \mu_\sigma[i] \\
					\phantom{R = } \left\{ V_{i}\rightarrow V_{i_{1}}V_{i_{2}}...V_{i_{k}} ~\left|~\begin{array}{l} 1\leq i,i_j \leq d, \\ \text{ for } 1\leq j\leq k \end{array}\right.\right\} & 
					\begin{array}{l} \theta(V_{i}\rightarrow V_{i_{1}}V_{i_{2}}...V_{i_{k}}) =  \\ \phantom{------} c^i_{i_{1},i_{2},...,i_{p}}  \end{array}\
					 \\		
					 \\					 
					\end{array}$}}		}}
	\caption{Transforming a PMTA into a PCFG}\label{fig:eqs-pmta-to-pcfg}
\end{figure}
The following proposition states that the transformation preserves the weights.
\begin{proposition}\label{prop:equiv1}
	$\grmr{W}(t)=\mathcal{A}(t)\quad$ for every $t\in \trees(\Sigma)$.
\end{proposition}

In Section~\ref{sec:learning-CMTAs}, Thm.~\ref{thm:bounds}, we show that we can learn a PMTA for a SUWCFG in polynomial time using a polynomial number of queries (see exact bounds there), thus obtaining the following result.
 \begin{corollary}\label{cor:learnSUWCFG}
	SUWCFGs can be learned in polynomial time using \smq s and 
	\seq s, where the number of \seq s is bounded by the number of non-terminal symbols.
\end{corollary}

The overall learning time for SUPCFG relies, on top of Corollary~\ref{cor:learnSUWCFG}, on the complexity of 
converting a WCFG into a PCFG~\parciteauthor{abney1999relating}, for which an exact bound is not provided, but the method is reported to converge quickly~\cite[\S 2.1]{smith2007weighted}.
%

\section{Learning Struc. Unamb. PCFGs}\label{sec:-tow-learning-CMTAs}

In this section we discuss the setting of the algorithm, the ideas behind Angluin-style learning algorithms, and the issues with using current algorithms to learn PCFGs.
As in \gstar (the algorithm for CFGs~\parciteauthor{Sakakibara88}), we assume an oracle that can answer two types of queries: \emph{structured membership queries} (\smq) and \emph{structured equivalence queries} (\seq) regarding the unknown regular tree series $\treesrs{T}$ (over a given ranked alphabet $\Sigma$). Given a structured string $s$, the query $\smq(s)$ is answered with the value $\treesrs{T}(s)$. Given an automaton $\aut{A}$ the query $\seq(\aut{A})$  is answered ``yes'' if $\aut{A}$ implements the skeletal tree series $\treesrs{T}$ and otherwise the answer is a pair $(s,\treesrs{T}(s))$ where $s$ is a  structured string for which $\treesrs{T}(s)\neq \aut{A}(s)$ (up to a  predefined error).

Our starting point is the learning algorithm \mstar~\parciteauthor{habrard2006learning} which learns MTA using \smq s and \seq s. 
 First we explain the idea behind this and similar algorithms, next the issues with applying it as is for learning PCFGs, then the idea behind restricting attention to strucutrally unambiguous grammars, and finally the algorithm itself.

\paragraph{Hankel Matrix}
All the generalizations of \lstar\ (the algorithm for learning regular languages using \mq s and \eq s, that introduced this learning paradigm~\parciteauthor{Angluin87}) 
share a general idea that can be explained as follows. 
A word or tree language as well as word or tree series can be represented by its Hankel Matrix. 
The Hankel Matrix has infinitely many rows and infinitely many columns. In the case of word series both rows and columns correspond to an infinite enumeration $w_0,w_1,w_2,\ldots$ of words over the given alphabet. In the case of tree series, the rows correspond to an infinite enumeration of trees $t_0,t_1,t_2,\ldots$ (where $t_i\in\trees(\Sigma)$) and the columns to an infinite enumeration of contexts $c_0,c_1,c_2,\ldots$ (where $c_i\in\trees_{\context}(\Sigma)$). The entry $\hm{H}(i,j)$ 
holds in the case of words the value for the word $w_i\cdot w_j$ and in the case of trees the value of the tree $\contextConcat{c_j}{t_i}$.
If the series is \emph{regular} there should exists a finite number of rows in this infinite matrix, which we term \emph{basis} such that all other rows can be represented using rows in the basis. In the case of \lstar, and \gstar (that learn word-languages and tree-languages, resp.) rows that are not in the basis should be equivalent to rows in the basis. In the case of \mstar (that learns tree-series) rows not in the basis should be expressible as a linear combination of rows in the basis. In our case, in order to apply the PMTA to PCFG conversion we need the algorithm to find a \emph{positive linear combination} of rows to act as the basis. In all cases we would like the basis to be \emph{minimal} in the sense that no row in the basis is expressible using other rows in the basis. This is since the size of the basis derives the dimension  of the automaton, and obviously we prefer smaller automata.

\paragraph{Positive linear spans}
An interest in \emph{positive linear combinations} occurs also in the research community studying convex cones and derivative-free optimizations and a theory of positive linear combinations has been developed~\parciteauthor{cohen1993nonnegative,Regis2016}.\footnote{Throughout the paper we use the terms \emph{positive} and \emph{nonnegative} interchangeably.} We need the following definitions and results.

The \emph{positive span} of a finite set of vectors $S=\{v_{1},v_{2},...,v_{k}\}\subseteq\mathbb{R}^{n}$ is defined as follows:
\begin{equation*}
\pos(S)=\{\lambda_{1}\cdot v_{1}+\lambda_{2}\cdot v_{2}+...+\lambda_{k}\cdot v_{k} ~|~ \lambda_{i}\geq 0, \forall 1\leq i\leq k\}
\end{equation*}
A set of vectors $S=\{v_{1},v_{2},...,v_{k}\}\subseteq\mathbb{R}^{n}$ is \emph{positively dependent} if some $v_{i}$ is a positive combination of the other vectors; otherwise, it is \emph{positively independent}. 
Let $A\in\mathbb{R}^{m\times n}$. We say that $A$ is \emph{nonnegative} if all of its elements are nonnegative.  The \emph{nonnegative column (resp. row) rank}  of $A$, denoted $\crank(A)$ (resp. $\rrank(A)$), is defined as the smallest nonnegative integer $q$ for which there exist a set of column- (resp. row-) vectors $V=\{v_{1},v_{2},...,v_{q}\}$  in $\mathbb{R}^{m}$ such that every column (resp. row) of $A$ can be represented as a positive combination of $V$. 
It is known  that $\crank(A)=\rrank(A)$ for any matrix $A$~\parciteauthor{cohen1993nonnegative}. Thus one can  freely use $\prank(A)$ for \emph{positive rank}, to refer to either one of these.

\paragraph{Issues with positive spans}

The first question that comes to mind, is whether we can use the \mstar\ algorithm as is to learn a positive tree series. We show that this is not the case.
In particular, there are positive tree series for which applying the \mstar\ algorithm results in an MTA with negative weights. Moreover, this holds also if
we consider word (rather than tree) series, and if we restrict the weights to be probabilistic (rather than simply positive).

\begin{proposition}\label{prop:needPMTAs}
	There exists a probabilistic word series for which the \mstar\ alg. may return an MTA with negative weights.
\end{proposition}
The proof shows this is the case for the word series over alphabet ${\Sigma=\{a,b,c\}}$ which assigns   
the following six strings: $aa$, $ab$, $ac$, $ba$, $cb$, $cc$ probability of $\frac{1}{6}$ each, and probability $0$ to all other strings.

Hence, we turn to ask whether 
we can adjust the algorithm \mstar\ to learn a positive basis. 
We note first that working with positive spans is much trickier than working with general spans, since for $d\geq 3$ there
is no bound on the size of a positively independent set in  $\mathbb{R}_+^d$~\parciteauthor{Regis2016}. To apply the ideas of the Angluin-style query learning algorihtms we 
 need the Hankel Matrix (which is infinite) to contain a finite sub-matrix with the same rank. 
Unfortunately, as we show next,
there exists a probabilistic (thus positive) tree series  $\treesrs{T}$
that can be recognized by a PMTA, but none of its finite-sub-matrices span the entire space of $\hm{H}_\treesrs{T}$.

\begin{proposition}\label{prop:pcfg-no-finite-rank}
	There exists a PCFG $\grmr{G}$ s.t. the Hankel Matrix
	$H_\grmr{G}$ corresponding to its tree-series $\treesrs{T}_G$ has the property that
	no finite number of rows positively spans the entire matrix.
\end{proposition}

The proof shows this is the case for the following PCFG:
$$\begin{array}{l@{\ \longrightarrow\ }l}
N_{1}& aN_{1}~[\frac{1}{2}]\ \mid\  aN_{2}~[\frac{1}{3}]\ \mid\  aa~[\frac{1}{6}]\\[2mm]
N_{2}& aN_{1}~[\frac{1}{4}]\ \mid\  aN_{2}~[\frac{1}{4}]\ \mid\  aa~[\frac{1}{2}]
\end{array}$$

\subsection{Focusing on Strucutrally Unambiguous CFGs}\label{sec:sucfg}
To overcome these obstacles we restrict attention to strucutrally unambiguous CFGs (SUCFGs) and their weighted/probabilistic versions (SUWCFGs/SUPCFGs).
A context-free grammar is termed \emph{ambiguous} if there exists more than one derivation tree for the same word.
We  term a CFG \emph{structurally ambiguous} if there exists more than one
derivation tree with the same structure for the same word.
A context-free  language is termed \emph{inherently ambiguous}   if it cannot be derived by an unambiguous CFG.
 Note that a CFG which is unambiguous is also structuraly unambiguous, while the other direction is not necessarily true.
{For instance, 
	the language $\{a^n b^n c^md^m~|~ n\geq 1, m\geq 1\}\cup \{a^n b^m c^md^n~|~ n\geq 1,m\geq 1\}$ which is inherently ambiguous~\cite[Thm.~4.7]{HopcroftUllman79} is not inherently structurally ambiguous.} 
Therefore
we have relaxed the classical unambiguity requirement. 

\paragraph{The Hankel Matrix and MTA for SUPCFG}
Recall that the Hankel Matrix considers skeletal trees. Therefore if a word has more than one derivation tree with the same
structure, the respective entry in the matrix holds the sum of weights for all derivations. This makes it harder for the learning algorithm
to infer the weight of each tree separately. By choosing to work with strucutrally unambiguous grammars, we overcome this diffictulty
as an entry corresponds to a single derivation tree. 

To discuss properties of the Hankel Matrix for an SUPCFG we need the following definitions.
Let ${H}$ be a matrix, $t$ a tree (or row index) $c$ a context (or column index), $T$ a set of trees (or row indices) and $C$ a set of contexts (or column indices).
We use ${H}[t]$ (resp. ${H}[c]$) for the row (resp. column) of ${H}$ corresponding to $t$ (resp. $c$).
Similarly we use ${H}[T]$ and ${H}[C]$ for the corresponding sets of rows or columns. Finally, we use  ${H}[t][C]$ for the restriction of ${H}$ to row $t$ and columns $[C]$.

Two vectors, $v_{1},v_{2}\in\mathbb{R}^{n}$ are co-linear with a scalar ${\alpha\in\mathbb{R}}$ for some $\alpha\neq 0$ iff $v_{1}=\alpha\cdot v_{2}$.
 Given a matrix $H$, and two trees $t_1$ and $t_2$, we say that $t_{1}\treesColin{\alpha}{H} t_{2}$ iff ${H}[t_{1}]$ and ${H}[t_{2}]$ are co-linear, with scalar $\alpha\neq 0$. That is, $H[t_{1}]=\alpha\cdot H[t_{2}]$. Note that if $H[t_{1}]=H[t_{2}]=\bar{0}$, then $t_{1}\treesColin{\alpha}{H} t_{2}\treesColin{\alpha}{H} t_{1}$ for every $\alpha>0$.
We say that $t_{1}\treesEquiv{H} t_{2}$ if $t_{1}\treesColin{\alpha}{H} t_{2}$ for some $\alpha\neq 0$.  It is not hard to see that $\treesEquiv{H}$ is an equivalence relation.

The following proposition states that in the Hankel Matrix of an SUPCFG, the rows of trees that are rooted by the same non-terminal
 are co-linear.
\begin{proposition}\label{prop:colinearity}
	Let $\hm{H}$ be the Hankel Matrix of an SUPCFG. 
	Let $t_{1},t_{2}$ be derivation trees rooted by the same non-terminal. Assume $\Prob(t_{1}),\Prob(t_{2})>0$.
	Then $t_{1}\treesColin{\alpha}{\hm{H}} t_{2}$ for some $\alpha\neq 0$.
\end{proposition}

We can thus conclude that the number of equivalence classes of $\equiv_{H}$ for an SUPCFG is finite and
bounded by the number of non-terminals plus one (for the zero vector). 
\begin{corollary}\label{cor:finite-equiv-cls}
	The skeletal tree-set for an SUPCFG has a finite number of equivalence classes under $\equiv_{H}$.
\end{corollary}

Next we would like to reveal the restrictions that can be emposed on a PMTA that corresponds to an SUPCFG.
We term an MTA \emph{co-linear} (and denote it CMTA) if in every column of every transition matrix $\mu_\sigma$ there is at most one entry
which is non-negative. 
\begin{proposition}\label{prop:SUWCFGhaveCMTA}
	A CMTA can represent an SUPCFG.
\end{proposition}
The proof relies on showing that a WCFG is strucuturally unambiguous iff it is invertible
and converting an invertible WCFG into a PMTA yields a CMTA.\footnote{A CFG $\grmr{G}=\langle \Vars,\Sigma,R,S\rangle$ is said to be
invertible if and only if $A \rightarrow \alpha$ and $B \rightarrow \alpha$ in $R$ implies $A = B$~\parciteauthor{Sakakibara92}.}

\section{The Learning Algorithm}\label{sec:learning-CMTAs}

 Let $\treesrs{T}:\trees(\Sigma)\rightarrow \R$ be an unknown tree series, and let $\hm{H}$ be its Hankel Matrix.
The learning algorithm \LearnCMTA (or \cstar, for short), provided in Alg.~\AlgLearnCmta, maintains a data structure called an \emph{observation table}. An observation table for $\treesrs{T}$ is a quadruple $(T,C,H,B)$. Where $T\subseteq\trees(\Sigma)$ is a set of row titles, $C\subseteq\trees_\context(\Sigma)$ is a set of column titles,  $H:T\times C\rightarrow\mathbb{R}$ is a sub-matrix of $\hm{H}$, and $B\subset T$, the so called \emph{basis}, is a set of row titles corresponding to rows of $H$ that are co-linearly independent. 
The algorithm starts with an almost empty observation table, where $T=\emptyset$, $C=\context$, $B=\emptyset$ and uses procedure $\Complete(T,C,H,B,\Sigma_0)$ to add the nullary symbols of the alphabet to the row titles, uses $\smq$ queries to fill in the table until certain criteria hold on the observation, namely it is \emph{closed} and \emph{consistent}, as defined in the sequel.
Once the table is closed and consistent, it is possible to extract from it a CMTA $\aut{A}$ (as we shortly explain). The algorithm then issues the query $\seq(\aut{A})$. If the result is ``yes'' the algorithm returns $\aut{A}$ which was determined to be structurally equivalent to the unknown series. Otherwise, the algorithm gets in return a counterexample $(s,\treesrs{T}(s))$, a structured string in the symmetric difference of $\aut{A}$ and $\treesrs{T}$, and its value. It then uses $\Complete$ to add all prefixes of $t$ to $T$ and uses \smq s to fill in the entries of the table until the table is once again closed and consistent.

\commentout{
	\begin{algorithm}
		\caption{$\LearnCMTA(T,C,H,B)$.}\label{alg:cstar}\label{alg:learn}
		\begin{algorithmic}[1]
			\State  Initialize $B\gets\emptyset,~T\gets\emptyset,~C\gets\{\context\}$ 
			\State $\Complete(T,C,H,B,\Sigma_0)$			
			\While{true}
			\State $\mathcal{A}\gets \ExtractCMTA(T,C,H,B)$
			\State $t\gets\seq(\mathcal{A})$
			\If{$t$ is null}
			\State \Return $\mathcal{A}$
			\EndIf
			\State $\Complete(T,C,H,B,  \pref(t))$
			\EndWhile
		\end{algorithmic}
	\end{algorithm}
	}

\quad \\
\noindent\makebox[.48\textwidth]{
	\includegraphics[scale=0.91,page=2, trim=1cm 22cm 10cm 1.8cm]{figures.pdf}
}
\noindent\makebox[.48\textwidth]{
	\includegraphics[scale=0.91,page=3, clip, trim=1cm 23.6cm 10cm 1.8cm]{figures.pdf}
}
\noindent\makebox[.48\textwidth]{
	\includegraphics[scale=0.91,page=4, clip, trim=1cm 22cm 10cm 1.8cm]{figures.pdf}

}
\noindent\makebox[.48\textwidth]{
	\includegraphics[scale=0.91,page=5, clip, trim=1cm 23.6cm 10cm 1.8cm]{figures.pdf}
}
Given a set of trees $T$ we use $\Sigma(T)$ for the set of trees $\{ \sigma(t_1,\ldots,t_k) ~|~\exists {\Sigma_k\in\Sigma},\ {\sigma\in\Sigma_k},\ {t_i\in T},\ {\forall 1\leq i \leq k}\}$.
The procedure $\Close(T,C,H,B)$, Alg.~\AlgClose, checks if $\hm{H}[t][C]$
is co-linearly independent from $T$ for some tree $t\in\Sigma(T)$. If so it adds $t$ to both $T$ and $B$ and loops back until no such trees are found, in which case  the table is termed \emph{closed}.

We use $\Sigma(T,t)$ for the set of trees in $\Sigma(T)$ satisfying that one of the children is the tree $t$. 
We use $\Sigma(T,\context)$ for the set of contexts all of whose children are   in $T$.
An observation table $(T,C,H,B)$ is said to be \emph{zero-consistent} if for every
tree $t\in T$ for which $H[t]=\overline{0}$ it holds that $H[\contextConcat{c}{t'}]=\overline{0}$ for every $t'\in \Sigma(T,t)$ and $c\in C$. It is said to be \emph{co-linear consistent} if for every $t_1,t_2\in T$ s.t. $t_1\treesColin{\alpha}{H} t_2$ 
and every context $c\in \Sigma(T,\context)$ we have that   $\contextConcat{c}{t_1}\treesColin{\alpha}{H} \contextConcat{c}{t_2}$. The procedure $\Consistent$, given in Alg.~\AlgConsistent, looks for trees which violate the zero-consistency or co-linear consistency requirement, and for every violation, the respective context is added to $C$. 

The procedure $\Complete(T,C,H,B,S)$, given in Alg.~\AlgComplete, first adds the trees in $S$ to $T$, 
then runs procedures $\Close$ and $\Consistent$ iteratively until the table is both closed and consistent. When the table is closed and consistent the algorithm extracts from it a CMTA as detailed in Alg.~\AlgExtractCmta.

Overall we can show that the algorithm always terminates, returning a correct CMTA whose dimension is minimal, namely it equals the rank $n$ of Hankel matrix for the target language.
It does so while asking at most $n$ equivalence queries, and the number of membership queries is polynomial in $n$, and in the size of the largest counterexample $m$, but of course exponential in $p$, the highest rank of the a symbol in $\Sigma$. Hence for a grammar in Chomsky Normal Form, where $p=2$, it is polynomial in all parameters.

\begin{theorem}\label{thm:bounds}
	Let $n$ be the rank of the target language, let $m$ be the size of the largest counterexample given by the teacher, and let $p$ be the highest rank of a symbol in $\Sigma$. Then the algorithm makes at most $n\cdot(n+m\cdot n+|\Sigma|\cdot (n+m\cdot n)^{p})$ $\smq$s and at most $n$ $\seq$s.
\end{theorem}

\noindent\makebox[.48\textwidth]{
	\includegraphics[scale=0.91,page=6, clip, trim=1cm 16.9cm 10cm 1.8cm]{figures.pdf}
}

\section{Demonstration}\label{sec:demonstration}
As a demonstration, we apply our algorithm to the learning of gene cluster grammars --- which is an important problem in functional genomics.
A \emph{gene cluster} is a group of genes that are co-locally conserved, not necessarily in the same order, across many genomes~\parciteauthor{winter2016finding}. The gene grammar corresponding to a given gene cluster describes its hierarchical inner structure and the relations between instances of the cluster succinctly;  assists in predicting the functional association between the genes in the cluster; provides insights into the evolutionary history of the cluster; aids in filtering meaningful from apparently meaningless clusters; and provides a natural and meaningful way of visualizing complex clusters. 

PQ trees have been advocated as a representation for gene-grammars~\parciteauthor{booth1976testing,bergeron2008formal}. A PQ-tree represents the possible permutations of a given sequence, and can be constructed in polynomial-time~\parciteauthor{landau2005gene}. 
A PQ-tree is a rooted tree with three types of nodes: {\em P-nodes}, {\em Q-nodes} and leaves. In the gene grammar inferred by a given PQ-tree, the children of a P-node can appear in any order, while the children of a Q-node must appear in either left-to-right or right-to-left order.

However,  the PQ tree model suffers from limited specificity, which often does not scale up to encompass gene clusters that exhibit some rare-occurring permutations.  It also does not model tandem gene-duplications, which are a common event in the evolution of gene-clusters. We exemplify how our algorithm can learn a grammar that addresses both of these problems. Using the more general model of context-free grammar, we can model evolutionary events that PQ-trees cannot, such as tandem gene-duplications. While the probabilities in our PCFGs grant our approach the capability to model rare-occurring permutations (and weighing them as such), thus creating a specificity which PQ-trees lack. 

We give two examples of gene-cluster grammars. The first is a PCFG describing a gene cluster corresponding to a multi-drug efflux pump (MDR). MDR's are used by some bacteria as a mechanism for antibiotic resistance, and hence are the focus of research aimed towards the development of new therapeutic strategies. 
In this example, we exemplify learning of a gene-cluster grammar which models distinctly ordered merge events of sub-clusters in the evolution of this pump. The resulting learned gene-cluster grammar is illustrated in Fig.~\ref{fig:grammar1}. A biological interpretation of the learned grammar, associating the highly probable rules with possible evolutionary events that explain them, is available in Section.~\ref{supp:biogramm}.

Note that, in contrast to the highly specific PCFG learned by our algorithm (Fig.~\ref{fig:grammar1}, top), the PQ-tree constructed for this gene cluster (Fig.~\ref{fig:grammar1}, middle) places all leaves under a single P-node, resulting in a complete loss of specificity regarding conserved gene orders and hierarchical swaps --- this PQ tree will accept all permutations of the four genes, without providing information about which permutations are more probable, as our learned PCFG does. 

In a second example, which is available in Section.~\ref{sec:fimacd_grammar} we exemplify learning of a gene-cluster grammar which models tandem duplication events. The yielded grammar demonstrates learning of an infinite language, with exponentially decaying probabilities. 

\begin{figure}
    \noindent\makebox[\textwidth/2]{
        \includegraphics[scale=0.8,page=9, clip, trim=2cm 20.5cm 9cm 1.8cm]{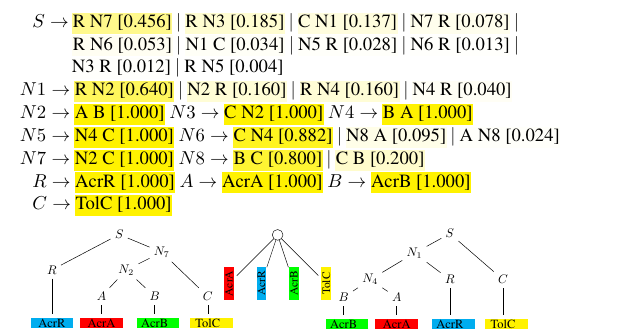}
    }
    \caption{{The PCFG learned from the MDR dataset. Also shown are the most probable tree according to the grammar (left) with $p=0.456$,  a non-probable tree (right) with $p=0.001$, and the PQ-tree (middle) showing a complete lack of specificity.}}
    \vspace{-4mm}
    \label{fig:grammar1}
\end{figure}


\section{Discussion}\label{sec:discussion}
 We have presented algorithms for learning structurally unambiguous PCFGs from a given black-box language model using structured membership and equivalence queries.
To our knowledge this is the first algorithm provided for this question. A recent paper~\parciteauthor{WeissGY19} advocates one can obtain an interpretable model 
of practically black-box models such as recurrent neural networks, using PDFA learning. The present work extends on this and offers obtaining intrepretable models also in cases where the studied
object exhibits non-regular (yet context-free) behavior, as is the case, e.g. in Gene Cluster grammars. 
}

\clearpage
\appendix

\clearpage

{
\newpage
\section{Running example}\label{app:example}
We will now demonstrate a running example of the learning algorithm. 
For the unknown target  consider the series which gives probability $\frac{1}{2}^n$ to
strings of the form $a^nb^n$ for $n\geq 1$ and probability zero to all other strings. 
 This series can be generated by the following SUPCFG $\grmr{G}=\langle \Vars,\{a,b\},R,S\rangle$
 with $\Vars=\{S,S_{2}\}$, and the following derivation rules:
$$\begin{array}{l@{\ \ \longrightarrow\ \ }lll}
S& aS_{2}~[\frac{1}{2}]\ \mid\ ab~[\frac{1}{2}]\\
S_{2}& Sb~~~[1]
\end{array}$$

The algorithm initializes $T=\{a,b\}$ and $C=\{\context\}$, fills in the entries of $M$ using $\smq$s, first for the rows of $T$ and then for their one letter extensions $\Sigma(T)$ (marked in blue), resulting   in the following observation table.
\begin{center}
	\begin{tabular}{ l | l}
		& \Tree [.$\context$ ] \\ \hline
		$a$& $0$\\ \hline
		$b$& $0$\\ \hline
		\color{blue} $?(a,a)$ & $0$\\\hline
		\color{blue} $?(a,b)$ & $0.5$\\\hline
		\color{blue} $?(b,a)$ & $0$\\\hline
		\color{blue} $?(b,b)$ & $0$\\\hline
	\end{tabular}
\end{center}

We can see that the table is not closed, since $?(a,b)\in\Sigma(T)$ but is not co-linearly spanned by $T$, so we add it to $T$. Also, the table is not consistent, since $a\treesColin{1}{H} b$, but $\mq(\contextConcat{\context}{?(a,b)})\neq\mq(\contextConcat{\context}{?(a,a)})$, so we add $?(a,\context)$ to $C$, and we obtain the following table.
From now on we  omit $0$ rows of $\Sigma(T)$ for brevity.
\begin{center}
	\begin{tabular}{l | l | l | l }
		& & \Tree [.$\context$ ] & $?(a,\context)$ \\ \hline
		$t_1$ & $a$ & $0$ & $0$ \\ \hline
		$t_2$ & $b$ & $0$ & $0.5$\\ \hline
		$t_3$ & $?(a,b)$ & $0.5$ &$ 0$ \\\hline
		& \color{blue} $?(?(a,b),b)$ & $0$ & $0.25$ \\
	\end{tabular}
\end{center}
The table is now closed but it is not zero-consistent, since we have $H[a]=\overline{0}$, but there exists a context with children in $T$, specifically $?(\context,b)$,
with which when $a$ is extended the result is not zero, namely $H[?(a,b)]\neq 0$. So we add this context
and we obtain the following table:

\vspace{2mm} 
\begin{center}
	\begin{tabular}{l | l | l | l | l }
		& & \Tree [.$\context$ ] & $?(a,\context)$ & $?(\context,b)$\\ \hline
		$t_1$ & $a$ & $0$ & $0$ & $0.5$ \\ \hline
		$t_2$ & $b$ & $0$ & $0.5$ & $0$ \\ \hline
		$t_3$ & $?(a,b)$ & $0.5$ & $0$ \\\hline
		$t_4\notin B$ & $?(a,a)$ & $0$ & $0$ & $0$ \\\hline
		& \color{blue} $?(?(a,b),b)$ & $0$ & $0.25$ & $0$ \\
	\end{tabular}
\end{center}
\vspace{2mm} 

Note that $t_{4}$ was added to $T$ since it wasn't spanned by $T$, but it is not a member of $B$, since $H[t_{4}]=0$. 
We can extract the following CMTA $\aut{A}_1=(\Sigma,\mathbb{R},d,\mu,\lambda)$ of dimension $d=3$ since $|B|=|\{t_{1},t_{2},t_{3}\}|=3$.
Let $\V=\R^3$.  For the letters $\sigma\in \Sigma_0=\{a,b\}$ we have that $\mu_\sigma:\V^0\rightarrow \V$, namely $\mu_a$ and $\mu_b$ are $3\times 3^0$-matrices. Specifically, following Alg.~\AlgExtractCmta we get that $\mu_{a}=(1,0,0)$, $\mu_{b}=(0,1,0)$ as $a$ is the first element of $B$ and $b$ is the second.
For ${?}\in\Sigma^2$ we have that $\mu_{?}:\V^2\rightarrow \V$, thus $\mu_{?}$ is a $3\times 3^2$-matrix.
We compute the entries of $\mu_{?}$ following Alg.~\AlgExtractCmta. For this, we consider all pairs of indices $(j,k)\in\{1,2,3\}^2$. For each such entry we look for the row $t_{j,k}=?(t_j,t_k)$
and search for the base row $t_i$ and the scalar $\alpha$ for which  $t_{j,k} \treesColin{\alpha}{H} t_i$.
We get that $t_{1,2}\treesColin{1}{H} t_3$, $t_{3,2}\treesColin{0.5}{H} t_2$ and for all other $j,k$ we get $t_{j,k}\treesColin{1}{H} t_{4}$, so we set $c^{i}_{j,k}$ to be $0$ for every $i$. Thus, we obtain the following matrix for $\mu_{?}$
$$
\eta_{?}=\begin{bmatrix}0&0&0&0&0&0&0&0&0\\0&0&0&0&0&0&0&0.5&0\\
0&1&0&0&0&0&0&0&0\end{bmatrix}$$
 The vector $\lambda$ is also computed via Alg.~\AlgExtractCmta, and we get $\lambda=(0,0,0.5)$. 
 
 The algorithm now asks an equivalence query and receives the following tree as a counter-example:
 
 $p=$\Tree [.? $a$ [.? [.? $a$ $b$ ] [.? [.? $a$ $b$ ] $b$ ] ] ]
 
 Indeed, while $\smq(p)=0$ we have that $\aut{A}(p)=0.125$. To see why $\aut{A}(p)=0.125$, let's look at the values $\mu(t)$ for every sub-tree $t$ of $p$. For the leaves, we have $\mu(a)=(1,0,0)$ and $\mu(b)=(0,1,0)$.
 
 Now, to calculate $\mu(?(a,b))$, we need to calculate $\mu_{?}(\mu(a),\mu(b))$. To do that, we first compose them as explained in the \emph{Multilinear functions} paragraph of Sec.~\ref{sec:mta}, see also Fig.~\ref{fig:MTA}. The vector $P_{\mu(a),\mu(b)}$ is: $P_{\mu(a),\mu(b)}=(0,1,0,0,0,0,0,0,0)$. When multiplying this vector by the matrix $\eta_{?}$ we obtain $(0,0,1)$. So $\mu(?(a,b))=(0,0,1)$. Similarly, to obtain  $\mu(?(?(a,b),a))$ we first compose the value $(0,0,1)$
 for $?(a,b)$ with the value $(0,1,0)$ for $a$ and obtain $P_{\mu?(a,b),\mu(a)}=(0,0,0,0,0,0,0,1,0)$. 
 Then we multiply $\eta_{?}$ by $P_{\mu?(a,b),\mu(a)}$ and obtain $(0,0.5,0)$. In other words, 
 $$\mu(?(?(a,b),a))=\mu_{?}(\begin{bmatrix}0\\ 0\\ 1\end{bmatrix},\begin{bmatrix}0\\ 1\\ 0\end{bmatrix})=\begin{bmatrix}0\\0.5\\0\end{bmatrix}.$$ 
 
 The following tree depicts the entire calculation by marking the values obtained for each sub-tree. We can see that  $\mu(p)=(0,0,0.25)$, thus we get that $\aut{A}=\mu(p)\cdot\lambda=0.125$.
 
\vspace{2mm} 
 \scalebox{.9}{
   \Tree [.$(0,0,0.25)$ $(1,0,0)$ [.$(0,0.25,0)$ [.$(0,0,1)$ $(1,0,0)$ $(0,1,0)$ ] [.$(0,0.5,0)$ [.$(0,0,1)$ $(1,0,0)$ $(0,1,0)$ ] $(0,1,0)$ ] ] ]
}
\vspace{2mm} 

 We add all prefixes of this counter-example to $T$ and we obtain the following table:
 \begin{center}
 	\begin{tabular}{l | l | l | l | l }
 		 & & \Tree [.$\context$ ] & \rotatebox[origin=c]{90}{$?(a,\context)$} & \rotatebox[origin=c]{90}{$?(\context, b)$} \\ \hline
 		$t_1$ & $a$  & $0$ & $0$ & $\frac{1}{2}$ \\ \hline
 		$t_2$ & $b$ & $0$ & $\frac{1}{2}$ &$0$ \\ \hline
 		$t_3$ & $?(a,b)$ & $\frac{1}{2}$ & $0$ & $0$ \\\hline
 		$t_4$ & $?(a,a)$ & $0$ & $0$ & $0$\\\hline
 		$t_5$ & $?(?(a,a),a)$ & $0$ & $0$ & $0$\\\hline
 		$t_6$ & $?(?(a,b),b)$ & $0$ & $\frac{1}{4}$ & $0$ \\\hline
 		$t_7$ & $?(?(a,b),?(?(a,b),b))$ & $0$ & $0$ & $0$\\\hline
 		$t_8$ & $?(a,?(?(a,b),?(?(a,b),b)))$ & $0$ & $0$ & $0$ \\
 	\end{tabular}
 \end{center}
 This table is not consistent since while $t_6 \treesColin{0.5}{H} t_2$ this co-linearity is not preserved when extended with $t_{3}=?(a,b)$ to the left, as evident from the context $?(a,\context)$.
We thus add the context $\contextConcat{?(a,\context)}{?(?(a,b),\context)}=?(a,?(?(a,b),\context))$ to obtain the final table:

 \begin{center}
 	\begin{tabular}{l | l | l | l | l | l}
 		& & \Tree [.$\context$ ] & \rotatebox[origin=c]{90}{$?(a,\context)$} & \rotatebox[origin=c]{90}{$?(\context, b)$} & \rotatebox[origin=c]{90}{$?(a,?(?(a,b),\context))$}\\ \hline
 		$t_1$ & $a$ & $0$ & $0$ & $\frac{1}{2}$ & $0$\\ \hline
 		$t_2$ & $b$ & $0$ & $\frac{1}{2}$ & $0$ & $\frac{1}{4}$ \\ \hline
 		$t_3$ & $?(a,b)$ & $\frac{1}{2}$& $0$ & $0$ & $0$ \\\hline
 		$t_4$ & $?(a,a)$ & $0$ & $0$ & $0$ & $0$ \\\hline
 		  & $?(?(a,a),a)$ & $0$ & $0$ & $0$ & $0$ \\\hline
 		$t_5$ & $?(?(a,b),b)$ & $0$ & $\frac{1}{4}$ & $0$ & $0$ \\\hline
 		  & $?(?(a,b),?(?(a,b),b))$ & $0$ & $0$ & $0$ & $0$\\\hline
 		  & $?(a,?(?(a,b),?(?(a,b),b)))$ & $0$ & $0$ & $0$ & $0$ \\\hline
 		 $t_6$  & \color{blue}$?(a,?(?(a,b),b))$ & $\frac{1}{4}$  & $0$ & $0$ & $0$ \\
 	\end{tabular}
 \end{center}
 The table is now closed and consistent, and we extract the following CMTA from it: 
 $\aut{A}_3=(\Sigma,\mathbb{R},4,\mu,\lambda)$ with $\mu_{a}=(1,0,0,0)$, $\mu_{b}=(0,1,0,0)$. 
 Now $\mu_{?}$ is a $4\times 4^2$ matrix. Its interesting entries are $c^3_{1,2}=1$, $c^4_{3,2}=1$ and $c^3_{1,4}=\frac{1}{2}$
 since $t_{1,2}\treesColin{1}{H} t_3$, $t_{3,2} \treesColin{1}{H} t_5$, $t_{1,5} \treesColin{1}{H} t_6 \treesColin{0.5}{H} t_3$.
 And  for every other combination of unit-basis vectors we have $t_{i,j}\treesColin{1}{H} t_4$.   
 The final output vector is $\lambda=(0,0,0.5,0)$.
 
 The equivalence query on this CMTA returns true, hence the algorithm now terminates, and we can convert this CMTA into a WCFG. 
 Applying the transformation provided in Fig.~\ref{fig:eqs-pmta-to-pcfg} we obtain the following WCFG:
 \begin{align*}
 S&\longrightarrow N_{3}~[0.5]   \\
 N_{1}&\longrightarrow a~[1.0 ]  \\
 N_{2}&\longrightarrow b~[1.0]   \\
 N_{3}&\longrightarrow N_{1} N_{2}~[1.0]\ \  |\ \ N_{1} N_{4}~[0.5]   \\
 N_{4}&\longrightarrow N_{3} N_{2}~[1.0]  
 \end{align*}
 
 Now, following~\cite{abney1999relating,smith2007weighted}
 we can calculate the partitions functions for each non-terminal. Let $f_{N}$ be the sum of the weights of all trees whose root is $N$, we obtain:
 \begin{align*}
 f_{S}&=1\\
 f_{N_{1}}&=1\\
 f_{N_{2}}&=1\\
 f_{N_{3}}&=2\\
 f_{N_{4}}&=2\\
 \end{align*}
 
 Hence we obtain the PCFG 
 \begin{align*}
 S&\longrightarrow N_{3}~[1.0]\\
 N_{1}&\longrightarrow a~[1.0]\\
 N_{2}&\longrightarrow b~[1.0]\\
 N_{3}&\longrightarrow N_{1} N_{2}~[0.5]\ |\ N_{1} N_{4}~[0.5]\\
 N_{4}&\longrightarrow N_{3} N_{2}~[1.0]
 \end{align*}
 which is a correct grammar for the unknown probabilistic series.

\newpage
\section{Omitted Proofs}\label{app:proofs}
\subsection{Proofs of Section 3}
Recall that given a WCFG $\tuple{\grmr{G},\theta}$, and a tree that can be derived from $\grmr{G}$,
namely some $t\in\deriv(\grmr{G})$, the weight of $t$ is given by $\theta(t)$.
Recall also that we are working with skeletal trees $s\in\struct(\deriv(\grmr{G}))$
and the weight of a skeletal tree $s$ is given by the sum of all derivation trees $t$
such that $s$ is the skeletal tree obtained from $t$ by replacing all non-terminals with $?$.

The following two lemmas and the following notation are used in  the proof of Proposition~\ref{prop:equiv1}.
For a skeletal tree $s$ and a non-terminal $N$ we use $\grmrwgt{N}{s}$ for the weight
of all derivation trees $t$ in which the root is labeled by non-terminal $N$ and $s$ is their skeletal form.

Assume $\grmr{G}=\langle \Vars,\Sigma,R,S\rangle$.
Lemma~\ref{lem:weight} below follows in a straight forward manner from the definition of $\mathcal{W}(\cdot)$ (given in
SubSec.~\ref{subsec:pcfgs}).

\begin{lemma}\label{lem:weight}
   
    Let ${s=?(s_1,s_2,\ldots,s_k)}$. Then the following holds for each non-terminal $N\in\Vars$:
    %
    \[\grmrwgt{N}{s}=\sum_{(X_{1},X_{2},\ldots,X_{k})\in\Vars^k}\begin{array}{l}\theta(N\rightarrow X_{1} X_{2}\cdots X_{k})\cdot \\ \grmrwgt{X_{1}}{s_1}\grmrwgt{X_{2}}{s_2}\cdots\grmrwgt{X_{k}}{s_k}\end{array}\]
    %
\end{lemma}

Recall that the transformation from a PMTA to a WCFG (provided in Fig.~\ref{fig:eqs-pmta-to-pcfg})
associates with every dimension $i$ of the PMTA $\aut{A}=(\Sigma,\mathbb{R}_+,d,\mu,\lambda)$
a variable (i.e. non-terminal) $V_i$.
The next lemma considers the $d$-dimensional vector $\mu(s)$ computed by $\aut{A}$ and
states that its $i$-th coordinate holds the value $\grmrwgt{V_i}{s}$.

\begin{lemma}\label{lem:vec-coord-vars}
    Let $s$ be a skeletal tree, and let $\mu(s)=(v_1,v_2,\ldots,v_d)$.
    Then $v_i=\grmrwgt{V_i}{s}$ for every $1\leq i \leq d$.
\end{lemma}

\begin{proof}
The proof is by induction on the height of $s$. For the base case $h=1$. Then $s$ is a leaf, thus $s\in\Sigma$. 
Then for each $i$ we have that $v[i]=\mu(\sigma)[i]$ by definition of MTAs computation.
On the other hand, by the definition of the transformation in  Fig.~\ref{fig:eqs-pmta-to-pcfg},
we have $\theta(V_i\rightarrow \sigma)=\mu(\sigma)[i]$. Thus, $\grmrwgt{V_i}{s}=\grmrwgt{V_i}{\sigma}=\mu(\sigma)[i]$, so the claim holds.

For the induction step, assume $s=?(s_{1},s_{2},...,s_{k})$. 
By the definition of a multi-linear map, for each $i$ we have:
\begin{equation*}
        v_i=\sum_{\left\{(j_1,j_2,\ldots,j_k)\in \{1,2,\ldots,d\}^k\right\}} c^i_{j_{1},...,j_{k}}v_{i}[j_{1}]\cdot...\cdot v_{k}[j_{k}]
\end{equation*}
where $c^i_{j_{1},...,j_{k}}$ are the coefficients of the $d\times d^k$ matrix of $\mu_?$ for $?\in\Sigma_k$.
By the definition of the transformation in  Fig.~\ref{fig:eqs-pmta-to-pcfg} we have that $c^i_{j_{1},...,j_{k}}=\theta(V_{i}\rightarrow V_{j_{1}}V_{j_{2}}...V_{j_{k}})$. 
Also, from our induction hypothesis, we have that for each $j_{i}$, $v_{i}[j_{i}]=\grmrwgt{V_{j_i}}{s_i}$. Therefore, we have that:
\[
        v_i=\sum_{V_{j_{1}}V_{j_{2}}...V_{j_{p}}\in \Vars^{k}} \begin{array}{l}
            \theta(V_{i}\rightarrow V_{j_{1}}V_{j_{2}}...V_{j_{k}})\cdot\\
            \grmrwgt{V_{j_1}}{s_1}\cdot...\cdot \grmrwgt{V_{j_k}}{s_k})
            \end{array}
\]
which according to Lemma \ref{lem:weight} is equal to $\grmrwgt{V_i}{s}$ as required.
\end{proof}

We are finally ready to prove \claimref{Proposition~\ref{prop:equiv1}} which states that\\
\begin{itemize}
    \item []
    \emph{  
    $\grmr{W}(t)=\mathcal{A}(t)\quad $ for every $t\in \skel(\trees(\Sigma))$. }
\end{itemize}

\begin{proof}
 
Let $\mu(t)=v=(v_{1},v_{2},...,v_{n})$ be the vector calculated by $\mathcal{A}$ for $t$. 
The value calculated by $\mathcal{A}$ is $\lambda\cdot v$, which is:
\begin{equation*}
    \sum_{i=1}^{n} v_{i}\cdot \lambda[i]
\end{equation*}
By the transformation in  Fig.~\ref{fig:eqs-pmta-to-pcfg} we have that $\lambda[i]=\theta(S\rightarrow V_{i})$ for each $i$. So we have:
\begin{equation*}
    \sum_{i=1}^{n} v_{i}\cdot \lambda[i] = \theta(S\rightarrow V_{i})\cdot v_{i}
\end{equation*}
By our claim, for each $i$, $v_{i}$ is equal to the probability of deriving $t$ starting from the non-terminal $V_{i}$, so we have that the value calculated by $\mathcal{A}$ is the probability of deriving the tree starting from the start symbol $S$. That is,  
$\grmr{W}(t)=\grmrwgt{S}{t}=\mathcal{A}(t)$.
\end{proof}

\subsection{Proofs of Section 4}
We start with the proof of \claimref{Proposition \ref{prop:needPMTAs}} which states that 
	\begin{itemize}
		\item [] 	\emph{There exists a probabilistic word series for which the \mstar\ algorithm may return an MTA with negative weights.}
	\end{itemize}

\begin{proof}
	The first rows of Hankel Matrix for this word series are given in the following figure (all entries not in the figure are $0$).
	One can see that the rows $\epsilon$, $b$, $c$, $ba$ are a positive span of the entire Hankel Matrix. 
	However, the $\mstar$ algorithm may return the MTA spanned by the basis $\epsilon$, $a$, $b$, $aa$. 
	Since the row of $c$ is obtained by substracting the row of $b$ from the row of $a$, this MTA will contain negative weights.
	
	\begin{center}
		\scalebox{.27}
		{\includegraphics{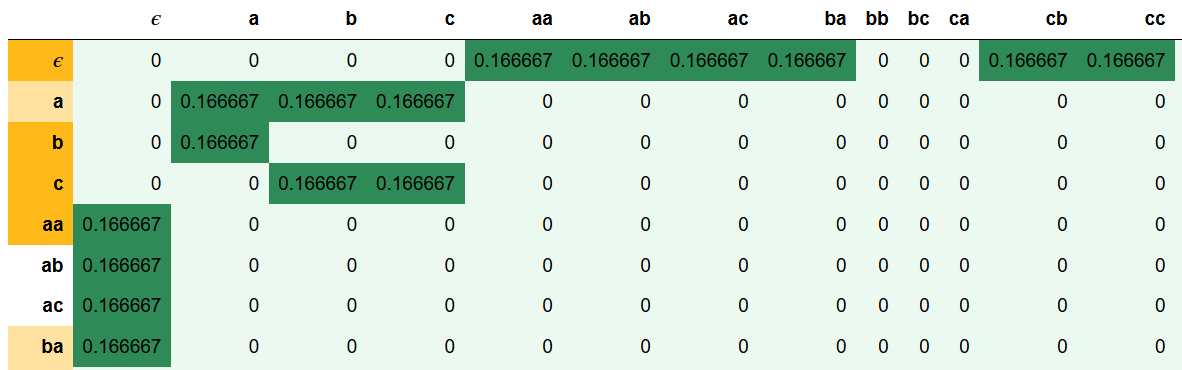}}
	\end{center}
\end{proof}
Next, we would like to show that there exists a PCFG $\grmr{G}$ such that the Hankel Matrix
$H_\grmr{G}$ corresponding to its tree-series $\treesrs{T}_G$ has the property that
no finite number of rows spans the entire matrix.

We first prove the following lemma about positive indepedent sets.

\begin{lemma}\label{lemmaSing}
	Let $B=\{b_{1},b_{2},...,b_{p}\}$ be set of positively independent vectors. 
	Let $\hat{B}$ be a matrix whose rows are the elements of $B$, and let $\alpha$ be a positive vector. 
	Then if $\hat{B}\alpha=b_{i}$ then $\alpha[i]=1$ and $\alpha[j]=0$ for every $j\neq i$.
\end{lemma}

\begin{proof}
	Assume $b_{i}=\hat{B}\alpha$. Then we have:
	\begin{equation*}
	    b_{i}=\hat{B}\alpha=\sum_{j=1}^{p}\alpha_{j}b_{j}=\alpha_{i}b_{i}+\sum_{j\neq i}\alpha_{j}b_{j}
	\end{equation*}
	If $\alpha_{i}<1$ we obtain:
	\begin{equation*}
	    b_{i}(1-\alpha_{i})=\sum_{j\neq i}\alpha_{j}b_{j}
	\end{equation*}
	Which is a contradiction, since $b_{i}\in B$ and thus can't be described as a positive combination of the other elements.
	
	If $\alpha_{i}>1$ we obtain:
	\begin{equation*}
	    \sum_{j\neq i}\alpha_{j}b_{j}+(\alpha_{i}-1)b_{i}=0
	\end{equation*}
	This is a contradiction, since each $b>0$, and each $\alpha_{j}\geq 0$, with $\alpha_{i}>1$.
	
	Hence $\alpha_{i}=1$. Therefore we have:
	\begin{align*}
	    \sum_{j\neq i}\alpha_{j}b_{j}=0
	\end{align*}
	Since $\alpha_{j}\geq 0$, and $b_{j}>0$ the only solution is that $\alpha_{j}=0$ for every $j\neq i$.
\end{proof}

\noindent
We are now ready to prove \claimref{Proposition~\ref{prop:pcfg-no-finite-rank}} which states:
\begin{itemize}
    \item []
	\emph{There exists a PCFG $\grmr{G}$ such that the Hankel Matrix
	$H_\grmr{G}$ corresponding to its tree-series $\treesrs{T}_G$ has the property that
	no finite number of rows spans the entire matrix.}
\end{itemize}

\begin{proof}
Let $\grmr{G}=(\{a\},\{N_{1},N_{2}\},R,N_{1})$ be the following PCFG:
$$\begin{array}{l@{\ \longrightarrow\ }l}
N_{1}& aN_{1}~[\frac{1}{2}]\ \mid\  aN_{2}~[\frac{1}{3}]\ \mid\  aa~[\frac{1}{6}]\\[2mm]
N_{2}& aN_{1}~[\frac{1}{4}]\ \mid\  aN_{2}~[\frac{1}{4}]\ \mid\  aa~[\frac{1}{2}]
\end{array}$$

We say that a tree has a \emph{chain structure} if every inner node is of branching-degree $2$ and has
one child which is a terminal. We say that a tree has a \emph{right-chain structure} (resp. \emph{left-chain structure})
if the non-terminal is always the right (resp. left) child.
Note that all trees in $\sema{G}$ have a right-chain structure (and the terminals are always the letter $a$), and can
be depicted as follows:

\Tree [.? $a$ [.? $a$ [.$\ddots$ $a$ [.? $a$ $a$ ] ] ] ]\vspace{2mm}
Let us denote by $p_{n}$ the total probability of all trees with $n$ non-terminals s.t. the lowest non terminal is $N_{1}$, 
and similarly, let us denote by $q_{n}$ the total probability of all trees with $n$ non-terminals s.t. the lowest non-terminal is $N_{2}$. 

We have that $p_{0}=0$, $p_{1}=\frac{1}{6}$, and  $p_{2}=\frac{1}{12}$. We also have that $q_{0}=0$, $q_{1}=0$ and  $q_{2}=\frac{1}{6}$. 

Now, to create a tree with $n$ non-terminals, we should take a tree with $n-1$ non-terminals, ending with either $N_{1}$ or $N_{2}$, and use the last derivation. So we have:
\begin{align*}
p_{n}&=\frac{1}{2}\cdot p_{n-1}+\frac{1}{4}\cdot q_{n-1}\\
q_{n}&=\frac{1}{3}\cdot p_{n-1}+\frac{1}{4}\cdot q_{n-1}
\end{align*}

We want to express $p_{n}$ only as a function of $p_{i}$ for $i<n$, and similarly for $q_{n}$. Starting with the first equation we obtain:
\begin{align*}
p_{n}&=\frac{1}{2}\cdot p_{n-1}+\frac{1}{4}\cdot q_{n-1}\\
4\cdot p_{n+1}-2\cdot p_{n}&=q_{n}\\
4\cdot p_{n}-2\cdot p_{n-1}&=q_{n-1}
\end{align*}

And from the second equation we obtain:
\begin{align*}
q_{n}&=\frac{1}{3}\cdot p_{n-1}+\frac{1}{4}\cdot q_{n-1}\\
3\cdot q_{n+1}-\frac{3}{4}\cdot q_{n}&=p_{n}\\
3\cdot q_{n}-\frac{3}{4}\cdot q_{n-1}&=p_{n-1}
\end{align*}

Now setting these values in each of the equation, we obtain:

\begin{align*}
4\cdot p_{n+1}-2\cdot p_{n}&=\frac{1}{3}\cdot p_{n-1}+\frac{1}{4}(4\cdot p_{n}-2\cdot p_{n-1})\\
p_{n+1}&=\frac{1}{2}\cdot p_{n}+\frac{1}{12}\cdot p_{n-1}+\frac{1}{16}(4\cdot p_{n}-2\cdot p_{n-1})\\
p_{n+1}&=\frac{1}{2}\cdot p_{n}+\frac{1}{12}\cdot p_{n-1}+\frac{1}{4}\cdot p_{n}-\frac{1}{8}\cdot p_{n-1}\\
p_{n+1}&=\frac{3}{4}\cdot p_{n}-\frac{1}{24}\cdot p_{n-1}
\end{align*}

And:

\begin{align*}
3\cdot q_{n+1}-\frac{3}{4}\cdot q_{n}&=\frac{1}{2}\cdot(3\cdot q_{n}-\frac{3}{4}\cdot q_{n-1})+\frac{1}{4}\cdot q_{n-1}\\
3\cdot q_{n+1}&=\frac{9}{4}\cdot q_{n}-\frac{1}{8}\cdot q_{n-1}\\
q_{n+1}&=\frac{9}{12}\cdot q_{n}-\frac{1}{24}\cdot q_{n-1}=\frac{3}{4}\cdot q_{n}-\frac{1}{24}\cdot q_{n-1}
\end{align*}
Let's denote by $t_{n}$ the probability that $\mathcal{G}$ assigns to $a^{n}$. This probability is:
\begin{equation*}
t_{n}=\frac{1}{6}\cdot p_{n-1}+\frac{1}{2}\cdot q_{n-1}
\end{equation*}
Since $$p_{n-1}=\frac{3}{4}\cdot p_{n-2}-\frac{1}{24}\cdot p_{n-3}$$ and $$q_{n-1}=\frac{3}{4}\cdot q_{n-2}-\frac{1}{24}\cdot q_{n-3}$$ we obtain:
{\small
\begin{align*}
t_{n}=&\frac{1}{6}\cdot(\frac{3}{4}\cdot p_{n-2}-\frac{1}{24}\cdot p_{n-3})+\frac{1}{2}\cdot(\frac{3}{4}\cdot q_{n-2}-\frac{1}{24}\cdot q_{n-3})\\
t_{n}=&\frac{1}{6}\cdot\frac{3}{4}\cdot p_{n-2}-\frac{1}{6}\cdot\frac{1}{24}\cdot p_{n-3}+\frac{1}{2}\cdot\frac{3}{4}\cdot q_{n-2}-\frac{1}{2}\cdot\frac{1}{24}\cdot q_{n-3}\\
t_{n}=&\frac{3}{4}\cdot(\frac{1}{6}\cdot p_{n-2}+\frac{1}{2}\cdot q_{n-2})-\frac{1}{24}\cdot(\frac{1}{6}\cdot p_{n-3}+\frac{1}{2}\cdot q_{n-3})=\\
       =&\frac{3}{4}\cdot t_{n-1}-\frac{1}{24}\cdot t_{n-2}
\end{align*}
}

Hence, overall, we obtain:
\begin{equation*}
t_{n}=\frac{3}{4}\cdot t_{n-1}-\frac{1}{24}\cdot t_{n-2}
\end{equation*}

Now, let $L$ be the skeletal-tree-language of the grammar $\mathcal{G}$, and let $H$ be the Hankel Matrix for this tree set. Note, that any tree $t$ whose structure is not a right-chain, would have $L(t)=0$, and also for every context $c$, $L(\contextConcat{c}{t})=0$. Similarly, every context $c$ who violates the right-chain structure, would have $L(\contextConcat{c}{t})=0$ for every $t$.

Let $T_{n}$ be the skeletal tree for the tree of right-chain structure, with $n$ leaves. We have that $L(T_{1})=0$, $L(T_{2})=\frac{1}{6}$, $L(T_{3})=\frac{1}{4}$, and for every $i>3$ we have $$L(T_{i})=\frac{3}{4}\cdot L(T_{i-1})-\frac{1}{24}\cdot L(T_{i-2}).$$ Let $v_{i}$ be the infinite row-vector of the Hankel matrix corresponding to $T_{i}$. We have that for every $i>3$, $$v_{i}=\frac{3}{4}\cdot v_{i-1}-\frac{1}{24}\cdot v_{i-2}.$$ Assume towards contradiction that there exists a subset of rows that is a positive base and spans the entire matrix $H$.

Let $B$ be the positive base, whose highest member (in the lexicographic order) is the lowest among all the positive bases. Let $v_{r}$ be the row vector for the highest member in this base. Thus, $v_{r+1}\in\posspan{B}$. Hence:
\begin{equation*}
v_{r+1}=\alpha \hat{B}
\end{equation*}
Also, $v_{r+1}=\frac{3}{4}\cdot v_{r}-\frac{1}{24}\cdot v_{r-1}$. Therefore,
\begin{align*}
\frac{3}{4}\cdot v_{r}-\frac{1}{24}\cdot v_{r-1}=\alpha\hat{B}\\
v_{r}=\frac{4}{3}\cdot\alpha\hat{B}+\frac{1}{18}\cdot v_{r-1}
\end{align*}

We will next show that $v_{r-1}$ and $v_r$ are co-linear, which contradicts our choice of $v_r$.
Since $v_{r-1}\in\posspan{B}$ we know $v_{r-1}=\alpha'\hat{B}$ for some $\alpha'$. Therefore, 
\begin{equation*}
v_{r}=(\frac{4}{3}\cdot\alpha+\frac{1}{18}\cdot\alpha')\hat{B}
\end{equation*}
Let $\beta=\frac{4}{3}\cdot\alpha+\frac{1}{18}\cdot\alpha'$. Since $\alpha$ and $\alpha'$ are non-negative vectors, so is $\beta$. And by lemma \ref{lemmaSing}  it follows that for every $i\neq r$:
\begin{equation*}
\beta_{i}=\frac{4\cdot\alpha_{i}}{3}+\frac{\alpha'_{i}}{18}=0
\end{equation*}
Since $\alpha_{i}$ and $\alpha'_{i}$ are non-negative, we have that $\alpha_{i}=\alpha'_{i}=0$.

Since for every $i\neq r$, $\alpha'_{i}=0$, it follows that $v_{r-1}=m\cdot v_{r}$ for some $m\in\mathbb{R}$. 
Now, $m$ can't be zero since our language is strictly positive and all entries in the matrix are non-negative. Thus, $v_{r}=\frac{1}{m}\cdot v_{r-1}$, and $v_{r}$ and $v_{r-1}$ are co-linear. We can replace $v_{r}$ by $v_{r-1}$, contradicting the fact that we chose the base whose highest member is as low as possible.
\end{proof}

We provide here the proof of \claimref{Proposition~\ref{prop:colinearity}} which states that 

\begin{itemize}
	\item [] 
	\emph{Let $H$ be the Hankel Matrix of an SUWCFG. 
	Let $t_{1},t_{2}$ be derivation trees rooted by the same non-terminal $N_i$. Assume $\Prob(t_{1}),\Prob(t_{2})>0$.
	Then $t_{1}\treesColin{\alpha}{H} t_{2}$ for some $\alpha\neq 0$.}
\end{itemize}

\begin{proof}
	Let $c$ be a context. Let $u\context v$ be the  yield of the context; that is, the letters with which the leaves of the context are tagged, in a left to right order. $u$ and $v$ might be $\varepsilon$. 
	We denote by $\Prob_{i}(c)$ the probability of deriving this context, while setting the context location to be $N_{i}$. That is:
	\begin{equation*}
	\Prob_{i}(c)=\Prob(S\underset{\mathcal{G}}{\overset{*}{\rightarrow}} u N_{i} v)
	\end{equation*}
	Let $\Prob(t_{1})$ and $\Prob(t_{2})$ be the probabilities for deriving the trees $t_{1}$ and $t_{2}$ respectively. So we obtain:
	
	\begin{align*}
	\Prob(\contextConcat{c}{t_{1}})&=\Prob_{i}(c)\cdot \Prob(t_{1})\\
	\Prob(\contextConcat{c}{t_{2}})&=\Prob_{i}(c)\cdot \Prob(t_{2})
	\end{align*}
	So we obtain, for every context $c$, assuming that $P_{i}(c)\neq 0$:
	\begin{equation*}
	\frac{\Prob(\contextConcat{c}{t_{1}})}{\Prob(\contextConcat{c}{t_{2}})}=\frac{\Prob(t_{1})}{\Prob(t_{2})}
	\end{equation*}
	For a context $c$ s.t. $P_{i}(c)=0$ we obtain that $\Prob(\contextConcat{c}{t_{1}})=\Prob(\contextConcat{c}{t_{2}})=0$. So for every context:
	\begin{equation*}
	\Prob(\contextConcat{c}{t_{1}})=\frac{\Prob(t_{1})}{\Prob(t_{2})}\Prob(\contextConcat{c}{t_{2}})
	\end{equation*}
	So $H[t_{1}]=\alpha\cdot H[t_{2}]$ for $\alpha=\frac{\Prob(t_{1})}{\Prob(t_{2})}$, so $H[t_{1}]$ and $H[t_{2}]$ are co-linear, and $t_{1}\treesColin{\alpha}{H} t_{2}$.
\end{proof}

We turn to prove \claimref{Corollary~\ref{cor:finite-equiv-cls}} which states that
\begin{itemize}
	\item []
	\emph{The skeletal tree-set for an SUPCFG has a finite number of equivalence classes under $\equiv_{H}$.}
\end{itemize}
\begin{proof}
	Since the PCFG is structurally unambiguous, it follows that for every skeletal tree $s$ there is a single tagged parse tree $t$ s.t. $\skel(t)=s$. So for every $s$ there is a single possible tagging, and a single possible non-terminal in the root. By Proposition \ref{prop:colinearity} every two trees $s_{1},s_{2}$  which are tagged by the same non-terminal, and in which $\Prob(s_{1}),\Prob(s_{2})>0$ are in the same equivalence class under $\equiv_{H}$. There is another equivalence class for all the trees $t\in\trees$ s.t. $\Prob(t)=0$. Since there is a finite number of non-terminals, there is a finite number of equivalence classes under $\equiv_{H}$.
\end{proof}

To prove Proposition~\ref{prop:SUWCFGhaveCMTA} we 
first show how to convert a WCFG into a PMTA. Then we claim, that in case the WCFG is structurally unambiguous
the resulting PMTA is a CMTA.

\paragraph{Converting a WCFG to a PMTA}
Let $\tuple{\grmr{G},\theta}$ be a WCFG where $\grmr{G}=\langle \Vars,\Sigma,R,S\rangle$. Suppose w.l.o.g that $\Vars=\{N_{0},N_{1},...,N_{|\Vars|-1}\}$, $\Sigma=\{\sigma_{0},\sigma_{1},...,\sigma_{|\Sigma|-1}\}$ and that $S=N_{0}$. Let $n=|\Vars|+|\Sigma|$. We define a function $\iota:\Vars\cup\Sigma\rightarrow\mathbb{N}_{\leq n}$ in the following manner:
\[
\iota(x) =
\begin{dcases*}
j
   & $x=N_{j}\in \Vars$\\
   |\Vars|+j
   & $x=\sigma_{j}\in \Sigma$
\end{dcases*}
\]
Note that since $\Vars\cap\Sigma=\emptyset$, $\iota$ is well defined. It is also easy to observe that $\iota$ is a bijection, so $\iota^{-1}:\mathbb{N}_{\leq n}\rightarrow \Vars\cup\Sigma$ is also a function.\\
We define a PMTA $\mathcal{A}_{\mathcal{G}}$ in the following manner:  
\begin{equation*}
    \mathcal{A}_{\mathcal{G}}=(\Sigma,\mathbb{R}_+,n,\mu,\lambda)
\end{equation*}
where
\begin{equation*}
    \lambda=[1,0,...,0]
\end{equation*}
(that is, $\lambda[0]=1$, and for $1<i\leq n$ $\lambda[i]=0$).

For each $\sigma\in\Sigma$ we define
\[
\mu_\sigma[i] =
\begin{dcases*}
1
   & $i=\iota(\sigma)$\\
   0
   & otherwise
\end{dcases*}
\]

For $(i,i_1,i_2,\ldots,i_j)\in \{1,2\ldots,|\Vars|\}^{|j|+1}$,
we define $R^{-1}(i,i_1,i_2,\ldots,i_j)$ to be the production rule
\[ \iota^{-1}(i)\longrightarrow \iota^{-1}(i_{1})~\iota^{-1}(i_{2})~\cdots~\iota^{-1}(i_{j}) \]
We define $\mu_?$ in the following way:
\[
{{\mu_?}^{i}}_{i_{1},...,i_{j}} =
\begin{dcases*}
\theta(R^{-1}(i,i_1,i_2,\ldots,i_j)) & $1\leq i\leq |\Vars|$\\
   0 & otherwise
\end{dcases*}
\]

We claim that the weights computed by the constructed PMTA
agree with the weights computed by the given grammar.

\begin{proposition}\label{prop:equiv2}
For each skeletal tree ${t\in \skels(\deriv(\grmr{G}))}$ we have that $\mathcal{W}_{\mathcal{G}}(t)=\mathcal{A}_{\mathcal{G}}(t)$.
\end{proposition}

\begin{proof}
The proof is reminiscent of the proof in the other direction, namely that of Proposition~\ref{prop:equiv1}.
We  first prove by induction that for each $t\in \skels(\deriv(\grmr{G})))$ the vector $\mu(t)=v=(v[1],v[2],...,v[n])$ calculated by $\mathcal{A}_{\mathcal{G}}$ maintains that for each $i\leq |\Vars|$, $v[i]=\grmrwgt{N_i}{t}$; and for $i>|\Vars|$ we have that $v[i]=1$ iff $t=\iota^{-1}(i)$ and $v[i]=0$ otherwise.

The proof is by induction on the height of $t$. For the base case $h=1$, thus $t$ is a leaf, therefore $t=\sigma\in\Sigma$. By definition $\mu_\sigma[i]=1$ if $i=\iota(\sigma)$ and $0$ otherwise. Hence $v[\iota(\sigma)]=1$, and for every $i\neq\iota(\sigma)$ $v[i]=0$. Since the root of the tree is in $\Sigma$, the root of the tree can't be a non-terminal, so $\grmrwgt{N_i}{t}=0$ for every $i$. Thus, the claim holds.

For the induction step,  $h>1$, thus $t=(? (t_{1}, t_{2},...,t_{k}))$ for some skeletal trees $t_{1}, t_{2},...,t_{k}$ of depth at most $h$. Let $\mu(t)=v=(v[1],v[2],...,v[n])$ be the vector calculated by $\mathcal{A}$ for $t$. By our definition of $\mu_?$, for every $i>|\Vars|$ ${{\mu_?}^{i}}_{i_{1},...,i_{j}}=0$ for all values of $i_{1},i_{2},...,i_{j}$. So for every $i>|\Vars|$ we have that $v[i]=0$ as required, since $t\notin\Sigma$. Now for $i\leq|\Vars|$. By definition of a multi-linear map we have that:
\begin{equation*}
        v[i]=\sum_{(i_1,i_2,\ldots,i_j)\in [|\Vars|]^j} {{\mu_?}^{i}}_{i_{1},...,i_{j}}\,v_{1}[j_{1}]\cdot...\cdot v_{j}[i_{j}]
\end{equation*}

Since $i\leq |\Vars|$, by our definition we have that:
\begin{equation*}
    {{\mu_?}^{i}}_{i_{1},...,i_{j}}\,=\theta(\iota^{-1}(i)\longrightarrow \iota^{-1}(i_{1})~\iota^{-1}(i_{2})~\cdots~\iota^{-1}(i_{j}))
\end{equation*}
For each $i_{k}$ let $B_{k}=\iota^{-1}(i_{k})$, also since $i\leq |\Vars|$, $\iota^{-1}(i)=N_{i}$, so:
\begin{equation*}
    {{\mu_?}^{i}}_{i_{1},...,i_{j}}\,=\theta(N_{i}\longrightarrow B_{1}B_{2}...B_{j})
\end{equation*}
For each $i$, by our induction hypothesis, if $t_{i}$ is a leaf, $v_{i}[j_{i}]=1$ only for $j_{i}=\iota(t_{i})$, and otherwise $v_{i}[j_{i}]=0$. If $t_{i}$ is not a leaf, then $v_{i}[j_{i}]=0$ for every $j_{i}>|\Vars|$; and for $j_{i}\leq |\Vars|$, we have that $v_{i}[j_{i}]=\grmrwgt{N_{j_i}}{t_i}$. Therefore we have:
\[
        v[i]=\sum_{(i_1,i_2,\ldots,i_j)\in [|\Vars|]^j} \begin{array}{l}
        \theta(N_{i}\rightarrow B_{1}B_{2}...B_{j})\cdot \\
        \grmrwgt{N_{i_1}}{t_1}\cdots \grmrwgt{N_{i_j}}{t_j}
        \end{array}
\]
So by lemma \ref{lem:weight} we have that $v[i]=\grmrwgt{N_{i}}{t}$ as required.

Finally, 
since $S=N_{0}$ and since by our claim, for each $i\leq |\Vars|$, $v_{i}=v[i]=\grmrwgt{N_{i}}{t}$, we get that $v[1]=\grmrwgt{S}{t}$. Also, since $\lambda=(1,0,...,0)$ we have that $\mathcal{A}_{\mathcal{G}}(t)$ is $v[1]$, which is $\grmrwgt{S}{t}$. Thus, it follows that  $\mathcal{W}_{\mathcal{G}}(t)=\mathcal{A}_{\mathcal{G}}(t)$
for every $t\in \skels(\deriv(\grmr{G}))$. 
\end{proof}

To show that the resulting PMTA is a CMTA we
need the following lemma.
We recall that a CFG $\grmr{G}=\langle \Vars,\Sigma,R,S\rangle$ is said to be
invertible if and only if $A \rightarrow \alpha$ and $B \rightarrow \alpha$ in $R$ implies $A = B$

\begin{lemma}\label{lem:invertible-iff-structuambig}
A CFG is invertible iff it is structurally unambiguous
\end{lemma}
\begin{proof}
Let $\grmr{G}$ be a SUCFG. We  show that $\grmr{G}$ is invertible. Assume towards contradiction that there are derivations $N_{1}\rightarrow\alpha$ and $N_{2}\rightarrow\alpha$. Then the tree $?(\alpha)$ is structurally ambiguous since its root can be tagged by both $N_{1}$ and $N_{2}$.

Let $\grmr{G}$ be an invertible grammar. We  show that $\grmr{G}$ is an SUCFG. Let $t$ be a skeletal tree. We show by induction on the height of $t$ that there is a single tagging for $t$.

For the base case, the height of $t$ is $1$. Therefore, $t$ is a leaf so obviously, it has a single tagging.

For the induction step, we  assume that the claim holds for all skeletal trees of height at most $h\geq 1$. Let $t$ be a tree of height $h+1$. Then $t=?(t_{1},t_{2},...,t_{p})$ for some trees $t_{1},t_{2},...,t_{p}$ of smaller depth. By the induction hypothesis, for each of the trees $t_{1},t_{2},...,t_{p}$ there is a single possible tagging. 
Hence we have established that all nodes of $t$, apart from the root, have a single tagging.
Let $X_{i}\in\Sigma\cup N$ be the only possible tagging for the root of $t_{i}$. Let $\alpha=X_{1}X_{2}...X_{p}$. Since the grammar is invertible, there is a single non-terminal $N$ s.t. $N\rightarrow\alpha$. Hence, there is a single tagging for the root of $t$ as well. Thus $\grmr{G}$ is structurally unambiguous.
\end{proof}

We are finally ready to prove \claimref{Proposition~\ref{prop:SUWCFGhaveCMTA}}:
\begin{itemize}
	\item []
	\emph{A CMTA can represent an SUWCFG.}
\end{itemize}

\begin{proof}
    By Proposition~\ref{prop:equiv2} a WCFG $\tuple{\grmr{G},\theta}$ 
    can be represented by a PMTA $\aut{A}_{\grmr{G}}$, namely they provide the same weight for every
    skeletal tree. By Lemma~\ref{lem:invertible-iff-structuambig} the fact that $\grmr{G}$ is unambiguous implies
    it is invertible. We show that given $\grmr{G}$ is invertible, the 
    resulting PMTA is actually a CMTA. That is, in every column of the matrices of $\aut{A}_{\grmr{G}}$,
    there is at most one non-zero coefficient. Let $\alpha\in (\Sigma\cup \Vars)^{p}$, let $\iota(\alpha)$ be the extension of $\iota$ to $\alpha$ (e.g. $\iota(aN_7bb)=\iota(a)\iota(N_7)\iota(b)\iota(b)$).
    Since $\grmr{G}$ is invertible, there is a single $N_{i}$ from which $\alpha$ can be derived, namely for which
     $\grmrwgt{N_{i}}{t^{N_i}_\alpha}>0$ where $t^{N_i}_\alpha$ is a tree deriving $\alpha$ with $N_i$ in the root.
     If $\alpha\in\Sigma$, i.e. it is a leaf, then we have that $\mu_\sigma[j]=0$ for every $j\neq i$, and $\mu_\sigma[i]>0$. If $\alpha\notin\Sigma$, then we have that ${{\mu_?}^{j}}_{\iota(\alpha)}=0$ for every $j\neq i$, and ${{\mu_?}^{i}}_{\iota(\alpha)}>0$, as required.
\end{proof}

\subsection{Proofs of Section 5}
We first provide a detailed description of the algorithms
that were not given in the body of the paper due to lack of space,
and then we delve into the proof of correctness of the learning algorithm.
To prove the main theorem we require a series of lemmas, which we state and prove here.

We start with some additional notations.
Let $v$ be a row vector in a given matrix. Let $C$ be a set of columns. We denote by $v[C]$ the restriction of $v$ to the columns of $C$. For a set of row-vectors $V$ in the given matrix, we denote by $V[C]$ the restriction of all vectors in $V$ to the columns of $C$.

\begin{lemma}\label{lemmaContextColin}
	Let $B$ be a set of vectors in a matrix $H$, and let $C$ be a set of columns. If a row $v[C]$ is co-linearly independent from $B[C]$ then $v$ is co-linearly independent from $B$.
\end{lemma}
\begin{proof}
	Assume towards contradiction that there is a vector $b\in B$ and a scalar $\alpha\in\reals$ s.t. $v=\alpha b$. Then for every column $c$ we have $v[c]=\alpha b[c]$. In particular that holds for every $c\in C$. Thus, $v[C]=\alpha b[C]$ and so $v[C]$ is not co-linearly independent from $B[C]$, contradicting our assumption.
\end{proof}

\begin{lemma}\label{replacementLemma}
	Let $\aut{A}=(\Sigma,\mathbb{R},d,\mu,\lambda)$ 
	be a CMTA. Let $t_{1},t_{2}$ s.t. $\mu(t_{1})=\alpha\cdot\mu(t_{2})$. Then for every context $c$:
	\begin{equation*}
	\mu(\contextConcat{c}{t_{1}})=\alpha\cdot\mu(\contextConcat{c}{t_{2}})
	\end{equation*}
\end{lemma}	
	\begin{proof}
		The proof is by induction on the depth of $\context$ in $c$.
		
	For the base case, 	the depth of $\context$ in $c$ is $1$. Hence, $c=\context$ and indeed we have:
			\begin{equation*}
			\mu(\contextConcat{c}{t_{1}})=\mu(t_{1})=\alpha\cdot\mu(t_{2})=\alpha\cdot\mu(\contextConcat{c}{t_{2}})
			\end{equation*}
			As required.

		For the induction step, 
		 assume the claim holds for all contexts where $\context$ is in depth at most $h$. Let $c$ be a context s.t. $\context$ is in depth $h+1$. Hence, there exists contexts $c_1$ and $c_2$ s.t.  $c=\contextConcat{c_{1}}{c_{2}}$ where $c_{2}=\sigma(s_{1},s_{2},...,s_{i-1},\context,s_{i+1},...,s_{p})$ for some $s_i$'s and the depth of $\context$ in $c_{1}$ is $h$. Let $t'_{1}=\contextConcat{c_{2}}{t_{1}}$ and let $t'_{2}=\contextConcat{c_{2}}{t_{2}}$ We have:
		 {\small{
		\begin{align*}
		\mu(t'_{1})&=
		\mu(\contextConcat{c_{2}}{t_{1}})\\
		&=\mu_{\sigma}(\mu(s_{1}),\mu(s_{2}),...,\mu(s_{i-1}),\mu(t_{1}),\mu(s_{i+1}),...,\mu(s_{p}))\\
		&=\mu_{\sigma}(\mu(s_{1}),\mu(s_{2}),...,\mu(s_{i-1}),\alpha\cdot\mu(t_{2}),\mu(s_{i+1}),...,\mu(s_{p}))
		\end{align*}}}
		Similarly for $t_{2}$ we obtain:
		\begin{equation*}
		\mu(t'_{2})=\mu_{\sigma}(\mu(s_{1}),\mu(s_{2}),...,\mu(s_{i-1}),\mu(t_{2}),\mu(s_{i+1}),...,\mu(s_{p}))\\
		\end{equation*}
		By properties of multi-linear functions we obtain:
		\begin{equation*}
		\begin{array}{rl}
		\mu(\sigma(s_{1},s_{2},...,s_{i-1},t_{1},s_{i+1},...,s_{p}))&=\\
		\alpha\cdot\mu(\sigma(s_{1},s_{2},...,s_{i-1},t_{2},s_{i+1},...,s_{p}))
		\end{array}
		\end{equation*}

		Thus, $\mu(t'_{1})=\alpha\cdot\mu(t'_{2})$, and by the induction hypothesis on $c_{1}$ we have:
		\begin{equation*}
		\mu(\contextConcat{c_{1}}{t'_{1}})=\alpha\cdot\mu(\contextConcat{c_{1}}{t'_{2}})
		\end{equation*}
		So:
		\begin{equation*}
		\mu(\contextConcat{c}{t_{1}})=\mu(\contextConcat{c_{1}}{t'_{1}})=\alpha\cdot\mu(\contextConcat{c_{1}}{t'_{2}})=\alpha\cdot\mu(\contextConcat{c}{t_{2}})
		\end{equation*}
		As required.	
\end{proof}

A subset $B$ of $T$ is called a \emph{basis} if for every ${t\in T}$, if ${H[t]\neq 0}$ then there is a unique $b\in B$, s.t. $t\treesColin{\alpha}{H} b$.
Let $(T,C,H,B)$ be an observation table. Then $B=\{b_1,b_2\ldots,b_d\}$ is a {basis} for $T$, and if $b_i$ is the unique element of $B$ s.t. $t\treesColin{\alpha}{H} b_i$,
we say that $\classrepr{t}{B}=b_{i}$, $\classcoeff{t}{B}=\alpha$, and $\classind{t}{B}=i$. 

The following lemma states that the value assigned to a tree $?(t_{1},t_{2},...,t_{p})$ all of whose children are in $T$, 
can be computed by multiplying the respective coefficients $\classcoeff{t_i}{B}$ witnessing the co-linearity of $t_i$ to its respective
base vector $\classrepr{t_i}{B}$.

\begin{lemma}\label{consistencyExpansion}
	Let $(T,C,H,B)$  be a closed consistent observation table. Let $t_{1},t_{2},...,t_{p}\in T$, and let $t=?(t_{1},t_{2},...,t_{p})$. Then:
	\begin{equation*}
	H[?(t_{1},t_{2},...,t_{p})]=\prod_{i=1}^{p}\classcoeff{t_{i}}{B}\cdot H[?(\classrepr{t_{1}}{B},\classrepr{t_{2}}{B},...,\classrepr{t_{p}}{B})]
	\end{equation*}
\end{lemma}

	\begin{proof}
		Let $k$ be the number of elements in $t_{1},t_{2},...,t_{p}$  s.t. $t_{i}\neq\classrepr{t_{i}}{B}$. We proceed by induction on $k$. 
		
		For the base case, we have $k=0$, so for every $t_{i}$ we have $t_{i}=\classrepr{t_{i}}{B}$ and $\classcoeff{t_{i}}=1$. Hence, obviously we have:
		\begin{equation*}
		H[?(t_{1},t_{2},...,t_{p})]=\prod_{i=1}^{p}\classcoeff{t_{i}}{B}\cdot H[?(\classrepr{t_{1}}{B},\classrepr{t_{2}}{B},...,\classrepr{t_{p}}{B})]
		\end{equation*}
		
		Assume now the claim holds for some ${k\geq 0}$. Since ${k+1>0}$ there is at least one $i$  s.t. $t_{i}\neq\classrepr{t_{i}}{B}$. Let $t'=?(t_{1},t_{2},...,t_{i-1},\classrepr{t_{i}}{B},t_{i+1},...,t_{p})$. Since the table is consistent, we have that $H[t]=\classcoeff{t_{i}}{B}\cdot H[t']$.
		Now,  $t'$ has $k$ children  s.t. $t_{i}\neq\classrepr{t_{i}}{B}$, so from the induction hypothesis we have:
		\begin{equation*}
		H[t']=\prod_{\begin{array}{c}{j=1}\\{j\neq i}\end{array}}^{p}\classcoeff{t_{j}}{B}\cdot H[?(\classrepr{t_{1}}{B},\classrepr{t_{2}}{B},...,\classrepr{t_{p}}{B})
		\end{equation*}
		So we have:
		\begin{equation*}
		H[t]=\classcoeff{t_{i}}{B}\cdot H[t']=\prod_{j=1}^{p}\classcoeff{t_{j}}{B}\cdot H[?(\classrepr{t_{1}}{B},\classrepr{t_{2}}{B},...,\classrepr{t_{p}}{B})
		\end{equation*}
		As required.
	\end{proof}

The following lemma states that if $t=?(t_{1},t_{2},...,t_{p})$  is co-linear to $s=?(s_{1},s_{2},...,s_{p})$ 
and $t_i$ is co-linear to $s_i$, for every $1\leq i\leq p$ and $H[y]\neq 0$ then the ratio between the tree coeficcient and the product of its children coefficeints is the same.
\begin{lemma}\label{lem:equal-prods-coef}
		Let $t=?(t_{1},t_{2},...,t_{p})$ and $s=?(s_{1},s_{2},...,s_{p})$ s.t. $t_{i}\treesEquiv{H} s_{i}$ for $1\leq i\leq p$. 
		Then $$\frac{\classcoeff{t}{B}}{\prod_{i=1}^{p}\classcoeff{t_{i}}{B}}=\frac{\classcoeff{s}{B}}{\prod_{i=1}^{p}\classcoeff{s_{i}}{B}}$$
\end{lemma}

\begin{proof}
 Let $t'=?(\classrepr{t_{1}}{B},\classrepr{t_{2}}{B} ...,\classrepr{t_{p}}{B})$.
 Note that we also have $t'=?(\classrepr{s_{1}}{B},\classrepr{s_{2}}{B},...,\classrepr{s_{p}}{B})$.
  Then from Lemma \ref{consistencyExpansion} we have that $H[t]=\prod_{i=1}^{p}\classcoeff{t_{i}}{B}\cdot H[t']$. 
  Similarly we have that $H[s]=\prod_{i=1}^{p}\classcoeff{s_{i}}{B}\cdot H[t']$. 
  Let $b=\classrepr{t}{B}=\classrepr{s}{B}$ we have $H[t]=\classcoeff{t}{B}\cdot H[b]$, and $H[s]=\classcoeff{s}{B}\cdot H[b]$.
Thus we have
\begin{equation*}
\classcoeff{t}{B}\cdot H[b]=\prod_{i=1}^{p}\classcoeff{t_{i}}{B}\cdot H[t']
\end{equation*}
And
\begin{equation*}
\classcoeff{s}{B}\cdot H[b]=\prod_{i=1}^{p}\classcoeff{s_{i}}{B}\cdot H[t']
\end{equation*}
Hence we have
\begin{equation*}
\frac{\classcoeff{t}{B}}{\prod_{i=1}^{p}\classcoeff{t_{i}}{B}}\cdot H[b]=H[t']=\frac{\classcoeff{s}{B}}{\prod_{i=1}^{p}\classcoeff{s_{i}}{B}}\cdot H[b]
\end{equation*}
Since $H[t]\neq 0$, and $t\treesEquiv{H} b\treesEquiv{H} t'$ we obtain that $H[b]\neq 0$ and $H[t']\neq 0$. 
Therefore
\begin{equation*}
\frac{\classcoeff{t}{B}}{\prod_{i=1}^{p}\classcoeff{t_{i}}{B}}=\frac{\classcoeff{s}{B}}{\prod_{i=1}^{p}\classcoeff{s_{i}}{B}}
\end{equation*}
\end{proof}

The next lemma relates the value $\mu(t)$ to  $t$'s coefficeint, $\classcoeff{t}{B}$, and the vector for respective row in the basis, $\classind{t}{B}$.

\begin{lemma}\label{lem:mu-t-rel}
	Let $t\in T$. If $H[t]\neq 0$ then $\mu(t)=\classcoeff{t}{B}\cdot e_{\classind{t}{B}}$. If $H[t]=0$ then $\mu(t)=0$.
\end{lemma}
\begin{proof}
	The proof is by induction on the height of $t$.
	
	For the base case, $t=\sigma$ is a leaf, for some $\sigma\in\Sigma$. If $H[t]\neq 0$, by Alg.~\AlgExtractCmta, 
	we set $\sigma^{\classind{t}{B}}$ to be $\classcoeff{t}{B}$, and for every $j\neq\classind{t}{B}$ we set $\sigma^{j}$ to be $0$, 
	so $\mu(t)=\mu_{\sigma} =\classcoeff{t}{B}\cdot e_{\classind{t}{B}}$ as required. 
	Otherwise, if $H[t]=0$ then we set $\sigma^{i}$ to be $0$ for every $i$, so $\mu(t)=0$ as required.

	For the induction step,  $t$ is not a leaf. Then $t=?(t_{1},t_{2},...,t_{p})$. 
	If $H[t]\neq 0$, then since $H$ is zero-consistent, we have for every $1\leq i\leq p$ that 
	$H[t_{i}]\neq 0$. So for every $1\leq j\leq p$ by induction hypothesis we have 
	$\mu(t_{j})=\classcoeff{t_{j}}{B}\cdot e_{\classind{t_{j}}{B}}$. So:
	\begin{equation*}
	\begin{array}{rl}
	\mu(t)=&\mu_{?}(\mu(t_{1}),~\ldots~,\mu(t_{p}))=\\
	=&\mu_{?}(\classcoeff{t_{1}}{B}\cdot e_{\classind{t_{1}}{B}},~\ldots~,\classcoeff{t_{p}}{B}\cdot e_{\classind{t_{p}}{B}}) \\
	\end{array}
	\end{equation*}

	Therefore we have:
	\begin{equation*}
	\mu(t)[j]=\sum_{j_{1},...,j_{p}\in [n]^{p}}\sigma^{j}_{j_{1},...,j_{p}}\cdot\classcoeff{t_{1}}{B} e_{\classind{t_{1}}{B}}[j_{1}]\cdots\classcoeff{t_{p}}{B} e_{\classind{t_{p}}{B}}[j_{p}]
	\end{equation*}

	Note that for every $j_{1},j_{2},...,j_{p}\neq \classind{t_{1}}{B},\classind{t_{2}}{B},...,\classind{t_{p}}{B}$ we have $\classcoeff{t_{1}}{B}\cdot e_{\classind{t_{1}}{B}}[j_{1}]~\cdots~\classcoeff{t_{p}}{B}\cdot e_{\classind{t_{p}}{B}}[j_{p}]=0$, thus
	\begin{align*}
	\mu(t)[j]&=\sigma^{j}_{\classind{t_{1}}{B},...,\classind{t_{p}}{B}}\cdot~ \classcoeff{t_{1}}{B}\cdot e_{\classind{t_{1}}{B}}[\classind{t_{1}}{B}]~\cdots~\classcoeff{t_{p}}{B}\cdot e_{\classind{t_{p}}{B}}[\classind{t_{p}}{B}]\\
	&=\sigma^{j}_{\classind{t_{1}}{B},...,\classind{t_{p}}{B}}\cdot \classcoeff{t_{1}}{B}\cdot \classcoeff{t_{2}}{B}~\cdots~\classcoeff{t_{p}}{B}
	\end{align*}

	By Alg.~\AlgExtractCmta, and Lemma~\ref{lem:equal-prods-coef} we have that 	
	$$\sigma^{j}_{\classind{t_{1}}{B},\classind{t_{2}}{B},...,\classind{t_{p}}{B}}=
	\left\{ \begin{array}{l@{\quad }l} 
	0 & \mbox{if } j\neq \classind{t}{B} \\
	\frac{\classcoeff{t}{B}}{\prod_{j=1}^{p}{B}\classcoeff{t_{p}}{B}} & \mbox{if } j=\classind{t}{B}
	\end{array}\right.$$
		hence we obtain:
	\begin{align*}
	\mu(t)[\classind{t}{B}]&=\sigma^{j}_{\classind{t_{1}}{B},\classind{t_{2}}{B},...,\classind{t_{p}}{B}}\cdot \classcoeff{t_{1}}{B}\cdot \classcoeff{t_{2}}{B}\cdots \classcoeff{t_{p}}{B} \\ &=\frac{\classcoeff{t}{B}}{\prod_{j=1}^{p}\classcoeff{t_{p}}{B}}\cdot {\prod_{j=1}^{p}\classcoeff{t_{p}}{B}}=\classcoeff{t}{B}
	\end{align*}
	Thus $\mu(t)=\classcoeff{t}{B}\cdot e_{\classind{t}{B}}$ as required.
	
	If $H[t]=0$ then $\sigma^{i}_{\classind{t_{1}}{B},\classind{t_{2}}{B},...,\classind{t_{p}}{B}}=0$ for every $i$, and we obtain
	\begin{equation*}
	\mu(t)=0  
	\end{equation*}
	as required.
\end{proof}

Next we show that  rows in the basis get a standard basis vector. 

\begin{lemma}\label{lemmaBase}
	For every $b_{i}\in B$, $\mu(b_{i})=e_{i}$ where $e_{i}$ is the $i$'th standard basis vector.
\end{lemma}
\begin{proof}
	By induction on the height of $b_{i}$.
	
	\textbf{Base case:} $b_{i}$ is a leaf, so $b=\sigma$ for $\sigma\in\Sigma_0$. By  Alg.~\AlgExtractCmta we set $\sigma_{i}$ to be $1$ and $\sigma_{j}$ to be $0$ for every $j\neq i$, so $\mu(b_{i})=e_{i}$.
	
	\textbf{Induction step:} $b_{i}$ is not a leaf. Note that by definition of the method $\Close$ (Alg.~\ref{proc:close}), all the children of $b_{i}$ are in $B$. So $b_{i}=\sigma(b_{i_{1}},b_{i_{2}},...,b_{i_{p}})$ for some base rows $b_{i_j}$'s. Let's calculate $\mu(b_{i})[j]$
	
	\begin{equation*}
	\mu(b_{i})[j]=\sum_{j_{1},j_{2},...,j_{p}\in [n]^{p}}\sigma^{j}_{j_{1},j_{2},...,j_{p}}\cdot\mu(b_{i_{1}})[j_{1}]\cdot\hdots\cdot\mu(b_{i_{p}})[j_{p}]
	\end{equation*}
	
	By the induction hypothesis, for every $1\leq j\leq p$ we have that $\mu(b_{i_{j}})[i_{j}]=1$, and $\mu(b_{i_{j}})[k]=0$ for $k\neq i_{j}$. So for every vector $j_{1},j_{2},...,j_{p}\neq i_{1},i_{2},...,i_{p}$ we obtain:
	
	\begin{equation*}
	\mu(b_{i_{1}})[j_{1}]\cdot\hdots\cdot\mu(b_{i_{p}})[j_{p}]=0
	\end{equation*}
	
	And for $j_{1},j_{2},...,j_{p}=i_{1},i_{2},...,i_{p}$ we obtain:
	
	\begin{equation*}
	\mu(b_{i_{1}})[j_{1}]\cdot\hdots\cdot\mu(b_{i_{p}})[j_{p}]=1
	\end{equation*}
	
	So we have:
	\begin{equation*}
	\mu(b_{i})[j]=\sigma^{i}_{i_{1},i_{2},...,i_{p}}
	\end{equation*}
	By Alg.~\AlgExtractCmta we have that $\sigma^{i}_{i_{1},i_{2},...,i_{p}}=1$ and $\sigma^{j}_{i_{1},i_{2},...,i_{p}}=0$ for $j\neq i$, so $\mu(b_{i})[i]=1$ and $\mu(b_{i})[j]=0$ for $j\neq i$. Hence $\mu(b_{i})=e_{i}$ as required.
\end{proof}

The next lemma states 
for a tree $t=\sigma(b_{i_{1}},b_{i_{2}},...,b_{i_{p}})$ with
children in the basis, if 
if $t\treesColin{\alpha}{H}b_i$ then  
 $\mu(t)=\alpha\cdot  e_{i}$ where $e_{i}$ 
 is the $i$'th standard basis vector.

\begin{lemma}\label{lemmaExtension}
	Let $t=\sigma(b_{i_{1}},b_{i_{2}},...,b_{i_{p}})$, s.t. ${b_{i_{j}}\in B}$ for ${1\leq j\leq p}$. Assume ${H[t]=\alpha\cdot H[b_{i}]}$ for some $i$. Then $\mu(t)=\alpha\cdot  e_{i}$.
\end{lemma}
\begin{proof}
	If $t=\sigma$ is a leaf, then by definition we have $\sigma_{i}=\alpha$ and $\sigma_{j}=0$ for $j\neq i$, so $\mu(t)=\alpha\cdot e_{i}$.
	
	Otherwise, $t$ isn't a leaf. Assume $t=\sigma(b_{i_{1}},b_{i_{2}},...,b_{i_{p}})$. We thus have
	\begin{equation*}
	\mu(t)[j]=\sum_{j_{1},j_{2},...,j_{p}\in [n]^{p}}\sigma^{j}_{j_{1},j_{2},...,j_{p}}\cdot\mu(b_{i_{1}})[j_{1}]\cdot\hdots\cdot\mu(b_{i_{p}})[j_{p}]
	\end{equation*}
	By Lemma \ref{lemmaBase} we have that $\mu(b_{i_{j}})=e_{i_{j}}$ for $1\leq j\leq p$, hence using a similar technique to the one used in the proof of Lemma \ref{lemmaBase} we obtain that for every $1\leq j\leq p$:
	\begin{equation*}
	\mu(t)[j]=\sigma^{j}_{i_{1},i_{2},...,i_{p}}
	\end{equation*}
	
	By Alg.~\AlgExtractCmta we have that $\sigma^{j}_{i_{1},i_{2},...,i_{p}}=\alpha$ for $i=j$ and $\sigma^{j}_{i_{1},i_{2},...,i_{p}}=0$ for $i\neq j$, so $\mu(t)=\alpha\cdot  e_{i}$ as required.
\end{proof}

The following lemma generalizes the previous lemma to any tree $t\in T$.

\begin{lemma}\label{lem:Correctness}
	Let $H$ be a closed consistent sub-matrix of the Hankel Matrix. Then for every $t\in T$ s.t. $H[t]=\alpha\cdot H[b_{i}]$ we have $\mu(t)=\alpha\cdot e_{i}$
\end{lemma}
\begin{proof}
	By induction on the height of $t$.
	For the base case $t$ is a leaf, and the claim holds by Lemma \ref{lemmaExtension}.
	
	Assume the claim holds for all trees of height at most $h$. Let $t$ be a tree of height $h$. Then $t=\sigma(t_{1},t_{2},...,t_{p})$. Since $T$ is prefix-closed, for every $1\leq j\leq p$ we have that $t_{j}\in T$. And from the induction hypothesis for every $1\leq j\leq p$ we have that $\mu(t_{j})=\alpha_{j}\cdot e_{i_{j}}$. Hence
	\begin{align*}
	\mu(t)&=\mu(\sigma(t_{1},t_{2},...,t_{p}))=\mu_{\sigma}(\mu(t_{1}),\mu(t_{2}),...,\mu(t_{p}))\\
	&=\mu_{\sigma}(\alpha_{1}\cdot e_{i_{1}},\alpha_{2}\cdot e_{i_{2}},...,\alpha_{p}\cdot e_{i_{p}})\\
	&=\prod_{j=1}^{p}\alpha_{j} \cdot \mu_{\sigma}(e_{i_{1}},e_{i_{2}},...,e_{i_{p}})
	\end{align*}
	Let $t'=\sigma(b_{i_{1}},b_{i_{2}},...,b_{i_{p}})$. From Lemma \ref{lemmaBase} we have
	\begin{align*}
	\mu(t)&=\prod_{j=1}^{p}\alpha_{j}\cdot\mu_{\sigma}(e_{i_{1}},e_{i_{2}},...,e_{i_{p}})\\
	&= \prod_{j=1}^{p}\alpha_{j}\cdot \mu(\sigma(\mu(b_{i_{1}}),\mu(b_{i_{2}}),...,\mu(b_{i_{p}})))\\
	&=\prod_{j=1}^{p}\alpha_{j}\cdot\mu(t')
	\end{align*}
	
	Since the table is consistent, we know that for each $1\leq j\leq p$ and $c\in C$:
	\begin{align*}
	H[\sigma(t_{1},t_{2},...,t_{j-1},t_{j},t_{j+1},...,t_{p})][c]=\\
	\alpha_{j}\cdot H[\sigma(t_{1},t_{2},...,t_{j-1},b_{i_{j}},t_{j+1},...,t_{p})][c]
	\end{align*}
	We can continue using consistency to obtain that
	\begin{align*}
	H[t][c]&=H[\sigma(t_{1},t_{2},...,t_{p})][c]\\
	&=\prod_{j=1}^{p}\alpha_{j}\cdot H[\sigma(b_{1},b_{2},...,,b_{p})][c]\\
	&=\prod_{j=1}^{p}\alpha_{j}\cdot H[t'][c]
	\end{align*}
	Thus $H[t]=\prod_{j=1}^{p}\alpha_{j}\cdot H[t']$. Let $\beta=\prod_{j=1}^{p}\alpha_{j}$, then $t\treesColin{\beta}{H} t'$. Let $b$ be the element in the base s.t. $t'\treesColin{\alpha}{H} b_{i}$. From Lemma \ref{lemmaExtension} we have that $\mu(t')=\alpha\cdot e_{i}$. Therefore $\mu(t)=\beta\cdot\alpha\cdot e_{i}$. 
	
	We have $\mu(t)=\beta\cdot\alpha\cdot e_{i}$ and $t\treesColin{\alpha\cdot\beta}{H} b_{i}$. Therefore the claim holds.
\end{proof}

We are now ready to show that for every tree $t\in T$ and context $c\in C$ the obtained CMTA
agrees with the observation table.

\begin{lemma}
	For every $t\in T$ and for every $c\in C$ we have that $\mathcal{A}(\contextConcat{c}{t})=H[t][c]$
\end{lemma}
\begin{proof}
	Let $t\in T$, s.t. $t\treesColin{\alpha}{H} b_{i}$ for some $b_{i}\in B$.
	The proof is by induction on the depth of $\context$ in $c$. 
	
	\textbf{Base case:} The depth of $c$ is $1$, so $c=\context$, and by lemma \ref{lem:Correctness} we have that $\mu(\contextConcat{c}{t})=\mu(t)=\alpha \cdot e_{i}$. Therefore $\mathcal{A}(t)=\alpha\cdot e_{i}\cdot\lambda$.  By Alg.~\AlgExtractCmta we have that $\lambda[i]=H[b_{i}][\context]$. So $\mathcal{A}(t)=\alpha\cdot H[b_{i}][\context]=H[t][\context]$ as required. 
	
	\textbf{Induction step:} Let $c$ be a context s.t. the depth of $\context$ is $h+1$. So $c=\contextConcat{c'}{\sigma(t_{1},t_{2},...,t_{i-1},\context,t_{i},...,t_{p})}$ for some trees $t_j\in T$, and some context $c'$  of depth  $h$.  For each $1\leq j\leq p$, let $b_{i_{j}}$ be the element in the base, s.t. $t_{j}\equiv_{H} b_{i_{j}}$, with co-efficient $\alpha_{j}$. Let $b$ be the element in the base s.t. $t\equiv_{H} b$ with coefficient $\alpha$. Let $\widetilde{t}$ be the tree:
	\begin{equation*}
	\widetilde{t}=\sigma(b_{i_{1}},b_{i_{2}},...,b_{i_{k-1}},b,b_{i_{k}},...,b_{i_{p}})
	\end{equation*}
	Note that $\widetilde{t}\in \Sigma(B)$ and hence $\widetilde{t}\in T$. From the induction hypothesis, we obtain:
	\begin{equation*}
	\mathcal{A}(\contextConcat{c'}{\widetilde{t}})=H[\widetilde{t}][c']
	\end{equation*}
	Since the table is consistent, we have:
	\begin{align*}
	H[t][c]&=H[\sigma(t_{i_{1}},t_{i_{2}},...,t_{i_{k-1}},t,t_{i_{k}},...,t_{i_{p}})][c']\\
	&=\alpha\cdot\prod_{i=1}^{p}\alpha_{i}\cdot H[\widetilde{t}][c']
	\end{align*}
	Let $\beta=\alpha\cdot\prod_{i=1}^{p}\alpha_{i}$.
	By definition of $\aut{A}$ we have: 
	\begin{align*}
	\aut{A}(\contextConcat{c'}{\sigma(b_{i_{1}},b_{i_{2}},...,b_{i_{k-1}},b,b_{i_{k}},...,b_{i_{p}})})=\\
	\mu(\contextConcat{c'}{\sigma(b_{i_{1}},b_{i_{2}},...,b_{i_{k-1}},b,b_{i_{k}},...,b_{i_{p}})}))\cdot\lambda
	\end{align*}
	Since each $t_{i_{j}}$ is in $T$, from Proposition \ref{lem:Correctness} we have that ${\mu(t_{i_{j}})=\alpha_{j}\cdot b_{i_{j}}}$, and that ${\mu(t)=\alpha\cdot\mu(b)}$.\\
	Let $\hat{t}=\sigma(t_{i_{1}},t_{i_{2}},...,t_{i_{k-1}},t,t_{i_{k}},...,t_{i_{p}}))$.
	 So
	\begin{align*}
	\mu(\hat{t}) 
	&=\mu(\sigma(t_{i_{1}},t_{i_{2}},...,t_{i_{k-1}},t,t_{i_{k}},...,t_{i_{p}})))\\
	&=\mu_{\sigma}(\alpha_{1}\cdot b_{i_{1}},...,\alpha_{k-1}\cdot b_{i_{k-1}},\alpha\cdot b,\alpha_{k}\cdot b_{i_{k}},...,\alpha_{p}\cdot b_{i_{p}})\\
	&=\alpha\cdot\prod_{j=1}^{p}\alpha_{i}\cdot\mu_{\sigma}(b_{i_{1}},...,b_{i_{k-1}},b, b_{i_{k}},...,b_{i_{p}})\\
	&=\beta\cdot\mu(\widetilde{t})
	\end{align*}
	By Lemma \ref{replacementLemma} we have that $$\mu(\contextConcat{c'}{\sigma(t_{i_{1}},t_{i_{2}},...,t_{i_{k-1}},t,t_{i_{k}},...,t_{i_{p}})})=\beta\cdot\mu(\contextConcat{c'}{\widetilde{t}})$$
	Hence
	\begin{align*}
	\mathcal{A}(\contextConcat{c}{t})&=\mathcal{A}(\contextConcat{c'}{\sigma(t_{i_{1}},t_{i_{2}},...,t_{i_{k-1}},t,t_{i_{k}},...,t_{i_{p}})})\\
	&=\mu(\contextConcat{c'}{\sigma(t_{i_{1}},t_{i_{2}},...,t_{i_{k-1}},t,t_{i_{k}},...,t_{i_{p}})})\cdot\lambda\\
	&=\beta\cdot\mu(\contextConcat{c'}{\widetilde{t}})\cdot\lambda
	\end{align*}
	Note that all the children of $\widetilde{t}$ are in $B$, and so $\widetilde{t}\in T$. Hence, from the induction hypothesis we have:
	\begin{equation*}
	H[\widetilde{t}][c']=\mathcal{A}(\contextConcat{c'}{\widetilde{t}})=\mu(\contextConcat{c'}{\widetilde{t}})\cdot\lambda
	\end{equation*}
	So:
	\begin{equation*}
	\mathcal{A}(\contextConcat{c}{t})=\beta\cdot H[\widetilde{t}][c']=H[t][c]
	\end{equation*}
	As required.
\end{proof}

\begin{proposition}
	In every iteration of Alg.~\AlgLearnCmta, the set $B$ maintained by the algorithm is contained in some correct solution.
\end{proposition}
\begin{proof}
	We  show by induction that in each iteration, $B\subseteq B^{*}$ for some co-linear base $B^{*}$.
	
	The base case is trivial since $B=\emptyset$. Hence, clearly $B\subseteq B^{*}$ for some co-linear base $B^{*}$.
	
	For the induction step, let $B$ be the set created by the algorithm in the previous iteration. Let $t$ be the tree picked by the algorithm to be added to the basis in the current iteration. If $t\in B^{*}$ then $B\cup\{t\}\subseteq B^{*}$. Otherwise, $t\notin B^{*}$. Let $N$ be the tagging for the root of $t$. Since $t$ was picked, we have that $H[t][C]$ is co-linearly independent from $B[C]$. By Lemma \ref{lemmaContextColin} $H[t]$ is co-linearly independent from $B$, and so by Prop.~\ref{prop:colinearity} there is no tree in $B$ whose root is tagged by $N$. Since $B^{*}$ is a co-linear base, there must be a tree $t'\in B^{*} \setminus B$ whose root is tagged by $N$, by Prop.~\ref{prop:colinearity} $t$ and $t'$ are co-linear. Let $B^{*'}=B^{*}\setminus \{t'\}\cup \{t\}$. $B^{*'}$ is still a co-linear base, because $t$ and $t'$ are co-linear. So, $B\cup\{t\}\subseteq B^{*'}$ and $B^{*'}$ is a co-linear base.
\end{proof}

\begin{proposition}\label{prop:rankInc}
	Let $(T,C,H,B)$ be a closed and consistent observation table. Let $t$ be a counterexample, and let $(T',C',H',B')=\Complete(T,C,H,B,\pref(t))$. Then $|B'|>|B|$. 
\end{proposition}
\begin{proof}
	Since $t$ was given as a counterexample, the previously extracted CMTA $\mathcal{A}=\textsl{CMTAB}(T,C,H,B)$ gave a wrong answer for it. By the method complete, $t\in T'$, hence by Lemma~\ref{lem:Correctness} we have that $\mathcal{A'}=\textsl{CMTAB}(T',C',H',B')$ agrees with $t$. Since we only add elements to the base, we have that $B\subseteq B'$. Now, clearly $B\neq B'$ since otherwise (following Alg.~\AlgExtractCmta) we would obtain $\mathcal{A}=\mathcal{A'}$. Hence $B\subsetneq B'$ implying $|B'|>|B|$.
\end{proof}

We are now ready to prove the main theorem,  \claimref{Theorem \ref{thm:bounds}} which states that 
\begin{itemize}
\item []
\emph{Let $n$ be the rank of the target-language, let $m$ be the size of the largest counterexample given by the teacher, and let $p$ be the highest rank of a symbol in $\Sigma$. Then the algorithm makes at most $n\cdot(n+m\cdot n+|\Sigma|\cdot (n+m\cdot n)^{p})$ $\mq$ and at most $n$ $\eq$.}
\end{itemize}

\begin{proof}
By Prop~\ref{prop:rankInc} we have that the rank of the finite Hankel Matrix $H[C]$ increases by at least one after each equivalence query. Since the rank of the infinite Hankel Matrix $H$ is $n$, it follows that the learner makes at most $n$ equivalence queries.

For the set of contexts, the algorithm starts with $C=\{\context\}$. The algorithm adds new contexts following failed consistency checks. Each added context thus separates two trees in $T$ and increases the rank of $H[C]$ by $1$. Therefore,  $|C|\leq n$. 

We add an element to set of trees $T$ in two cases: if it is in $\Sigma(T)$, and it is co-linearly independent from $T$, or it is a prefix of a counterexample given to us by the teacher. 

The first case can occur at most $n$ times, since each time we add an element from $\Sigma(T)$ to $T$ we increase the rank of the table by $1$. 

Since the number of $\eq$ made by the learner is at most $n$, the learner receives at most $n$ counterexamples from the teacher. Let $m$ be the size of the largest counterexample given by the teacher. We add all its prefixes to $T$, so each counterexample adds at most $m$ elements to $T$. So we have that by the end of the algorithm, $|T|\leq n+m\cdot n$. Let $p$ be the maximal rank of symbol in $\Sigma$, then $|\Sigma(T)|\leq |\Sigma|\cdot |T|^{p}\leq|\Sigma|\cdot (n+m\cdot n)^{p}$. Hence the number of rows in the table is at most:
\begin{equation*}
    n+m\cdot n+|\Sigma|\cdot (n+m\cdot n)^{p}
\end{equation*}
Since the number of columns in the table is at most $n$,  the algorithm makes at most:
\begin{equation*}
    n\cdot(n+m\cdot n+|\Sigma|\cdot (n+m\cdot n)^{p})
\end{equation*}
membership queries. For a Chomsky Normal Form grammar, we have that $p=2$ and therefore
\begin{equation*}
    n\cdot(n+m\cdot n+|\Sigma|\cdot (n+m\cdot n)^{2})
\end{equation*}
\end{proof}

\newpage
\clearpage
\section{Supplementary Material for the Demonstration}\label{app:bio}
We implemented our algorithm within a tool called PCFGLearner. The code and the data-sets used in this section, as well as a manual explaining how to employ our tool, are available with the submitted supplementary material. In this section we apply PCFGLearner to the learning of two PCFGs over genomic data. 

The section starts with a description of how the string data-set was generated (Section \ref{app:methods}).
Then, the conversion of the strings to parse trees is described in Section \ref{supp:trees}. 
\commentout
{
we generate a parse-tree for each sequence, using a variation of the CKY algorithm \cite{younger1967recognition,kasami1966efficient} as detailed in App.~\ref{supp:trees}, and assign it a probability according to its relative number of occurrences in the data-set.} 
Next, the implementation details of the oracle are given: Section \ref{supp:editDist} describes how the $\smq$ was implemented, and Section \ref{supp::randomSEQ} describes how the $\seq$ was implemented. 
A biological interpretation of the first grammar is given in Section \ref{supp:biogramm}, and the description of the second grammar is given in Section \ref{supp:bioFimACD}.

\subsection{Methods and datasets}\label{app:methods}
Genes in our experiment are represented by their membership in Clusters of Orthologous Genes (COGs) \cite{tatusov2000cog}.
$1,487$ fully sequenced prokaryotic strains with COG ID annotations were downloaded from GenBank (NCBI; ver 10/2012). The gene clusters were generated using the tool CSBFinder-S \cite{csbfinder-s}. 

CSBFinder-S was applied to all the chromosomal genomes in the dataset after removing their plasmids, using parameters $q=1$ (a colinear gene cluster is required to appear in at least one genome) and $k=0$ (no insertions are allowed in a colinear gene cluster), resulting in 595,708 colinear gene clusters.  Next, ignoring strand and gene order information, colinear gene clusters that contain the exact same COGs were united to form the generalized set of gene clusters.

To generate the trees for the first example, a subset of strings over the COGs AcrA(COG0845), AcrB(COG0841), TolC(COG1538) and AcrR(COG1309), where each of these COGs appeared at least once, was considered. This yielded 415 instances over 11 distinct strings. After the removal of strand information, 7 distinct strings remained, from which the trees were constructed. 

To generate the trees for the second example, a subset of strings over the COGs FimA(COG3539), FimC(COG3121), FimD(COG3188), and CitB(COG2197) was considered. This yielded 1899 instances over 28 distinct strings. After the removal of strand information, 25 distinct strings remained, from which the trees were constructed. 

In both examples, the constructed trees were annotated with a probability, according to the frequency of the corresponding strings in the dataset.  For the sake of simplicity and efficiency, we used binary trees in the given examples.

\subsection{Tree construction}\label{supp:trees}
Since the gene-cluster-data is available as strings, and our learning algorithm accepts trees, we propose the following approach to construct the expected parse trees from the strings by modelling two of the main events in operon evolution: in-tandem gene duplications~\parciteauthor{lewis1951pseudoallelism,labedan1995widespread} and progressive merging of sub-operons~\parciteauthor{fani2005origin,fondi2009origin}.

To model the former event, if two homologous genes (represented by the same COG) are found next to each other across many genomes, we assume that this is a result of a duplication event, and we place these two genes in the same sub-tree.
To model the latter event, we assume that if a sub-string is significantly over-represented in our data-set (indicating that it is conserved across many bacteria), then it is likely that this sub-string encodes a conserved functional unit, and that it should be confined to a distinguished sub-tree. 

The following scoring measure is used by the proposed parser to represent how likely it is that a considered parse tree structurally interprets a string.
For a given string $s$, let $s[i]$ be the $i$'th character in the string, and let $s[i:j]$ be the sub-string from the $i$'th character to the $j$'th character.
\commentout
{For every $i$ let $M_{i}$ be the maximal sub-string to which $S[i]$ belongs, in which all the characters are identical. 
First we define a \textbf{violating} sub-string, as a sub-string $s[i:j]$ in which there exists $i\leq k\leq j$ s.t $M_{k}\nsubseteq s[i:j]$. 

A \textbf{valid} parse-tree is an unlabeled derivation tree, whose yield is non-violating, and in which every non-leaf node can either have exactly two non-leaf children, or all of his children are leaves. In the latter case we denote such a node as a \textit{sub-string node}.
}

Given a weight function $w:\Sigma^{*}\rightarrow \mathbb{R}^{+}$, for a given string $s=\sigma_{1}\sigma_{2}...\sigma_{n}$, let $w_{\textsl{tree}}(s)$ denote the following function:
\[
 w_{\textsl{tree}}(s) = 
  \begin{cases} 
   w(s) &\text {if } |s|\leq 2 \\
 \begin{split}
   w(s)+\max_{1\leq i\leq n} (&w_{\textsl{tree}}(s[1:i])+\\
   &w_{\textsl{tree}}(s[i+1:n]))\end{split}  &\text { otherwise}\\
  \end{cases}
\]
\commentout
{
Let $s$ be a string of length $n$. And let $1\leq i,j\leq n$. We denote by $\mathcal{T}[i,j]$ the set of all possible valid parse trees over the string $s[i:j]$. Let us mark by $\opt[i,j]$ the maximal weight of a tree in $\mathcal{T}[i,j]$ that is:
\begin{equation*}
    \opt[i,j]=\max_{t\in\mathcal{T}[i,j]} w(t)
\end{equation*}
Clearly we have:
\begin{equation*}
    \begin{split}\opt[i,j]=
        \max(&w(S[i:j]),\max_{i\leq k<j} \opt[i,k]+\\
        &\opt[k+1,j]+w(S[i:j]))
    \end{split}
\end{equation*}
}
The optimal score $w_{\textsl{tree}}(s)$, as well as a corresponding optimal tree, can now be computed in $O(n^{3})$ using a simple dynamic-programming algorithm, which is a variation of the CKY algorithm \cite{kasami1966efficient, younger1967recognition}.

To force consecutive runs of a gene to appear together, the string $s$ is pre-processed before applying the algorithm. For each $\sigma\in\Sigma$, consecutive runs of $\sigma^{k}$ are merged to a new symbol $\sigma_{k}$. The new symbol $\sigma_{k}$ doesn't appear in $s$, and maintains that $w(u\sigma v)=w(u\sigma_{k} v)$ for every $u,v\in\Sigma^{*}$. 

After computing the optimal tree for the pre-processed string, every leaf whose label is $\sigma_{k}$ is replaced by a right-chain (defined in the next section) containing exactly $k$ leaves tagged with $\sigma$.

The tree construction algorithm is used both during the conversion of sequences to trees, as well as during the execution of an equivalence query of the MDR experiment (see section~\ref{supp::randomSEQ}). 

The scoring $w:\Sigma^{*}\rightarrow\mathbb{R}$ is implemented according to  \cite{svetlitsky2019csbfinder}, and we refer the reader to that paper for a detailed elaboration on how it is computed.  In a nutshell, for a given string $s$, and a given dataset $G$, let $q_s$ denote the number of genomes from $G$ in which $s$ occurs.  Assuming a uniform random order of genes in a genome, the ranking score computation first evaluates how likely it is for $s$ to occur in at least $q_s$ genomes from $G$ by mere chance, and then reports the negative logarithm (base $e$) of the computed score, so that the higher the ranking score of $s$, the less likely to be formed merely by chance.  The parameters considered in the ranking score computation for $s$ include the number of input genomes in $G$, the average length of a genome, the length of $s$, the number of genomes from $G$ in which $s$ occurs, and the frequencies of each gene from $s$ in the data.

\commentout{
\begin{algorithm} [H]
			\begin{algorithmic}[1]
			    \STATE $A \gets \text{empty matrix of size } n\times n$
                \FOR {$d\gets 0$ to $n$}
                \FOR {$i\gets 0$ to $n-d$}
                \STATE $j \gets i+d$
                \STATE $A[i][j] \gets w(S[i:j])$ 
                \FOR {$k\gets i$ to $j$}
                \IF {$A[i][k]+A[k+1][j]>A[i][j]$}
                \STATE $A[i][j] \gets A[i][k]+A[k+1][j]$
                \ENDIF
                \ENDFOR
                \ENDFOR
                \ENDFOR
			\end{algorithmic}
		\caption{$\text{FindOptimalParse}(S,w)$}\label{alg:parse}
\end{algorithm}}
\subsection{Membership query implementation}\label{supp:editDist}
In this section we describe the edit-distance functions that were used in the implementation of the oracle, for $\smq$ computation. We assume that all trees here are binary, and that their labels are in $\mathbb{N}$. We  use the notation $l(t)$ for the label of the tree $t$.

For a tree $t$, if $t$ is not a leaf, let $t_{1}$ and $t_{2}$ be its left and right children, respectively.

Two trees $t,s$ are \emph{incompatible} if $t$ and $s$ are both leaves, and $l(t)\neq l(s)$, or if only one of $t,s$ is a leaf, while the other is an internal node. 

\subsubsection{Swap-Event-Counting Edit Distance}
The swap-event-counting edit distance measure $w(t,s)$ is computed using the following recursive formula:
\[
 w(t,s) = 
  \begin{cases} 
   0 &\text {if } t \text{ and } s \\
           &\text{are leaves and }\\
           &l(t)=l(s) \\
   \infty &\text {if } t \text{ and } s \\
           &\text{are incompatible }\\
 \begin{split}
   \min(&w(t_{1},s_{1})+w(t_{2},s_{2}),\\
   &w(t_{1},s_{2})+w(t_{2},s_{1})+1)\end{split}  &\text { otherwise}\\

  \end{cases}
\]
\subsubsection{Duplication-Event-Counting Edit Distance}
A tree $t$ is a right-chain if $t$ is a leaf, or if the left child of $t$ is a leaf, the right child of $t$ is a right-chain and the labels on all the leaves of $t$ are equal. (Symmetrically, for a left-chain.) 

For a right or left chain $t$, let $\textsl{size}(t)$ be the number of leaves of $t$, and let $\textsl{label}(t)$ be the tagging of the leaves of $t$. (Note that in a chain, all the leaves have the same tagging, so $\textsl{label}$ is well-defined.)

Two trees $t$ and $s$, are \emph{right-chain-homologous}, if both are right-chains with the same tagging. For two chain-homologous trees, we say that $|\textsl{size}(t)-\textsl{size}(s)|$ is the \emph{copy-number difference} between them. 
For two trees $t$ and $s$, we say that $t$ and $s$ are \emph{right-homologous} if either $t$ and $s$ are right-chain-homologous, or if $t=(t_{1},t_{2})$, $s=(s_{1},s_{2})$ and both $t_{1}$ and $s_{1}$, and $t_{2}$ and $s_{2}$ are right-homologous. In that case, the copy-number difference between $t$ and $s$ is the sum of the copy-number differences between $t_{1}$ and $s_{1}$, and $t_{2}$ and $s_{2}$. (Symmetrically, for left-homologous trees.)

For two given trees $t,s$, let $w(t,s)$ denote the duplication-event-counting measure. The value of $w(t,s)$ is the copy-number difference between $t$ and $s$ if they are right-homologous, or $\infty$ if they aren't. The choice of right-homologous instead of left-homologous is arbitrary. 

\[
 w(t,s) = 
  \begin{cases} 
   |\textsl{size}(t)-\textsl{size}(s)| &\text {if } s \text{ and } t \text{ are}\\
                                       &\text{right-chain-homologous}\\
   \infty &\text{if } s \text{ and } t \text{ are incompatible}\\
          &\text{and aren't}\\
          &\text{right-chain-homologous}\\
   w(t_{1},s_{1})+w(t_{2},s_{2}) & \text {otherwise}\\
  \end{cases}
\] 
\commentout
{
$d(t,s)$ would be the copy-number difference between the trees, assuming that $t$ and $s$ are right-chain-homologous. It would be $\infty$ if one of the trees $t,s$ is not a right chain, or if both are right chains but with different labels. And would be the copy-number difference in the number of leaves between them if both are right chains with the same tagging. Formally:
\[
 d(t,s) = 
  \begin{cases} 
   |\textsl{size}(t)-\textsl{size}(s)| &\text {if } s \text{ and } t \text{ are}\\
                                       &\text{right-chain-homologous}\\
   \infty & \text {otherwise}\\
  \end{cases}
\] 
$e(t,s)$ would be the copy-number difference between the trees assuming that $t$ and $s$ aren't right-chain homologous, but are homologous. It would be $\infty$ if $t$ and $s$ are incompatible, and would be $0$ if both $t$ and $s$ are leaves and are equal. Otherwise, it would equal to the sum of $w(t_{1},s_{1})$ and $w(t_{2},s_{2})$.
\[
 e(t,s) = 
  \begin{cases} 
   0 &\text {if } t \text{ and } s \\
           &\text{are leaves and }\\
           &l(t)=l(s) \\
   \infty &\text {if } t \text{ and } s \\
           &\text{are incompatible }\\
   w(t_{1},s_{1})+w(t_{2},s_{2})&\text { otherwise}\\
  \end{cases}
\]
Finally we define $w(t,s)$ to be:
\begin{equation*}
    w(t,s)=\min(e(t,s),d(t,s))
\end{equation*}
}
\subsection{Equivalence query implementation}\label{supp::randomSEQ}
PCFGLearner supports three types of equivalence queries: Exhaustive Search, Random Sampling and Duplications Generator. In all these implementations, the oracle generates a set of trees $T$. Then, the oracle looks for a tree $t\in T$ s.t. $\mathcal{A}(t)\neq\smq(t)$. If such a tree is found, it is returned as a counterexample, otherwise, a positive answer to the equivalence query is returned. The three implementations differ  in the way  they generate the trees.

In the Exhaustive Search equivalence query, given an alphabet $\Sigma$, and a maximum length $l$, all strings up to length $l$ are generated. For each string $s$, the optimal tree for it $t$ is constructed, as described in Section \ref{supp:trees}.

In the Random Sampling equivalence query, instead of generating all strings up-to length $l$,  two parameters are given, $r$ and $s$. For each $\seq$, $r$ strings are independently sampled from a uniform distribution of all strings of length at most $s$. To each string the oracle then constructs the optimal tree, as in the Exhaustive Search equivalence query.

In the Duplications Generator, given a set of trees $R$, and a parameter $d$.  The oracle generates for each $t\in R$ all the trees $t'$ that can be obtained from $t$ by duplicating each of its leaves at most $d$ times. 

\subsection{Biological interpretation of the Multi Drug Resistance Efflux Pump grammar}\label{supp:biogramm}


In this section, we propose some biological interpretation of the grammar learned in the first example given in Section \ref{sec:demonstration} of the main paper, associating the highly probable rules of this grammar with explanatory evolutionary events, or with the functional reasoning underlying the strong conservation of the corresponding gene orders.  

AcrAB–TolC is a multidrug efflux pump that is widely distributed among gram-negative bacteria and extrudes diverse substrates from the cell, conferring resistance to a broad spectrum of antibiotics \cite{kobylka2020acrb}. This pump belongs to the resistance-nodulation-cell division (RND) family. In Gram-negative bacteria, RND pumps exist in a tripartite form, composed of an outer-membrane protein (TolC in our example), an inner membrane protein (AcrB in our example), and a periplasmic membrane fusion protein (AcrA in our example) that connects the other two proteins. The genes of the RND pump are often flanked with genes that code for local regulatory proteins, such as, in our example, the response regulator AcrR. 



 The pair of genes encoding the AcrB and AcrA proteins usually appear as an adjacent pair in our data, with AcrA preceding AcrB in the direction of transcription. The conserved order AcrA-AcrB could be explained by the order of assembly of the products of these genes into the AcrAB complex \cite{shi2019situ}, and by stochiometry  \cite{lalanne2018evolutionary}. The grammar learned for the AcrABR-TolC gene cluster (Figure \ref{fig:grammar1} in the main paper) indeed reflects this structural phenomenon, as $AB$, which is derived from the non-terminal $N_{2}$, is $4$ times more likely to be derived than $BA$, which is derived from the non-terminal $N_{4}$.


The outer-membrane protein TolC consistently appears, in our training dataset, adjacently to the AcrA-AcrB gene pair. However it forms a separate sub-tree from AcrAB in the highly probable trees generated by the learned grammar. This is due to the fact that, while the ordered pair AcrA-AcrB is very highly conserved in our data ($w(\text{AcrA }\text{AcrB})=31816.63$), TolC typically joins this triplet either upstream ($w(\text{TolC }\text{AcrA})=9494.15$) to it or downstream ($w(\text{AcrB }\text{TolC})=15783.51$) to it.

 Indeed, several studies indicate that the assembly and docking of the AcrAB-TolC efflux pump occur as a multi-step process, starting with the assembly of the AcrA-AcrB complex (see Figure \ref{fig:pump}), and only then activating the pump by the docking of TolC to the AcrAB complex \cite{ge2009c,shi2019situ}. 
 
 This assembly process is explained by additional studies, speculating that the TolC channel and inner membrane efflux AcrA-AcrB components may form a transient complex with the outer membrane channel TolC during efflux, due to the fact that TolC should be ready to use for not only AcrAB but also other efflux, secretion and transport systems in which it participates \cite{hayashi2016acrb}.
Indeed, in our general dataset, TolC participates in additional gene clusters encoding other export systems. This, along with the fact that in our string dataset this gene appears both upstream and downstream to the AcrAB pair, support the hypothesis that during the evolution of this pump across a wide-range of gram-negative bacteria, TolC was merged more than once with AcrAB (as well as with other export systems), in distinct evolutionary events. 


The gene cluster includes another gene, AcrR, that codes for a protein regulating the expression of the tripartite pump. This gene appears in a separate subtree from the AcrAB-TolC sub-tree, furthermore it is more likely to appear upstream to the AcrAB-TolC sub-tree in the highly probable trees.
 This could be explained by AcrR's functional annotation as a response regulator, whose role is to respond to the presence of a substrate, and consequently to enhance the expression of the RND tripartite efflux pump genes \cite{alvarez2013rnd}. 

 The fact that AcrR forms a separate subtree from the tripartite pump further exemplifies the role of merge events in gene cluster evolution. Indeed, other gene clusters in our dataset that include the tripartite pump are flanked by alternative response regulators, indicating that AcrAB-TolC homologs have merged with various regulators throughout the evoulution of gram-negative bacteria, yielding response to a variety of distinct drugs \cite{weston2018regulation}.
 
 Thus, the grammar learned by PCFGLearner exemplifies how our proposed approach can be harnessed to study biological systems that are conserved as gene clusters, and to explore
their function and their evolution.
\begin{figure}
    \scalebox{0.35}{\includegraphics{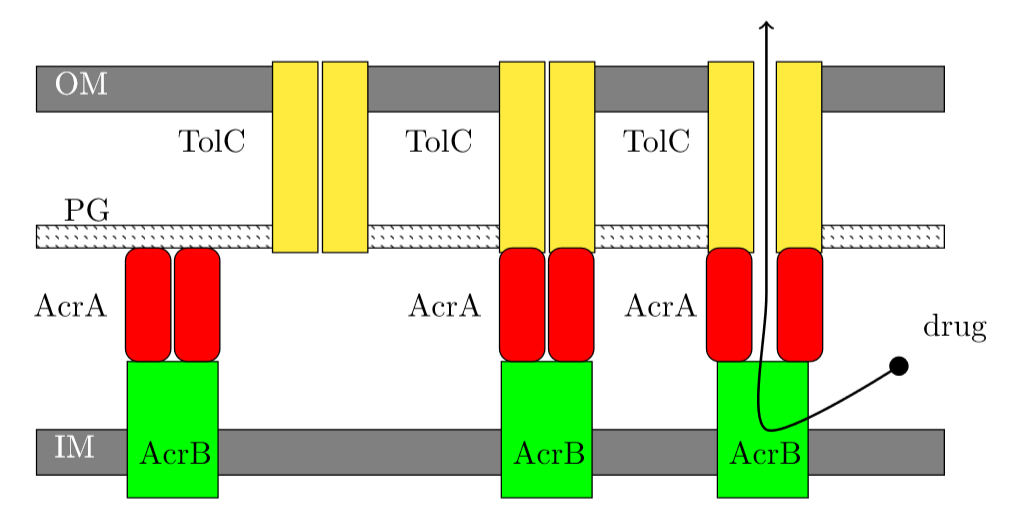}}
    \caption{In vivo assembly and functioning mechanism for multidrug efflux pump AcrAB-TolC according to  \cite{ge2009c,shi2019situ}. First AcrB associates with AcrA, to form the bipartite complex AcrAB. Next, AcrA changes its conformation to recruit TolC. Once TolC binds with the AcrAB bipartite complex, the fully  assembled tripartite pump remains in the resting state. When AcrB encounters a drug molecule, the pump adopts conformation accompanied with a contraction along the long axis and the substrate is expelled through the channel and out of the cell. This figure was prepared according to Figure 4 in \cite{shi2019situ}. OM, outer membrane; PG, peptidoglycan; IM, inner membrane.}
    \vspace{-4mm}
    \label{fig:pump}
\end{figure}
\subsection{A grammar exemplifying duplication events}\label{supp:bioFimACD}
In this example, we applied PCFGLearner to the FimACD gene cluster data-set, using the duplication-event-counting edit-distance metric with decay factor $q=0.2$, and the Duplications Generator equivalence query with parameter $d=2$.

In the rest of this section we give a few selected rules from the resulting grammar, mainly to demonstrate that this grammar can generate an infinite number of strings, with exponentially decaying probabilities. The entire grammar is available in section \ref{sec:fimacd_grammar}.
The exemplified grammar allows each symbol in $\Sigma$ to duplicate with a probability of $0.2$ for each duplication. For example, for FimA we have:
\begin{align*}
    N_{7}&\rightarrow N_{1} N_{8} [0.8]\\
    N_{7}&\rightarrow N_{1} N_{7} [0.2]\\
    N_{8}&\rightarrow N_{1} N_{1} [1.0]\\
    N_{1}&\rightarrow FimA [1.0]
\end{align*}
We refer the reader to Figures~\ref{fig:fimACD_tree} and \ref{fig:fimA_dup}, for an example of a parse tree from the grammar, and an illustration of how the right-chain with $n$ FimA-tagged leaves can be obtained with probability of $0.8\cdot (0.2)^{n-3}$ for $n\geq 3$ from $N_{7}$. The term $n\geq 3$ is due to the fact that $N_{7}$ already represents a sub-tree learned from the data-set, with three consecutive copies of FimA.
\clearpage
\commentout
{For a duplication of FimC, we can derive $N_{10}\rightarrow N_{28} N_{29} [0.146]$, instead of $N_{10}\rightarrow N_{28} N_{3} [0.586]$. Note that this derivation is $4$ times less likely in our grammar. For further duplications of FimC, one can proceed with $N_{29}\rightarrow N_{3} N_{29} [0.2]$. The derivation of the sub-tree is completed, when the rule $N_{29}\rightarrow N_{3} N_{3} [0.8]$ is used (see figure \ref{fig:fimC_dup}).

Since a sequence of three FimA's appears in the data-set, one can notice that the probability of deriving three sequential FimA's is much larger than that of deriving three consecutive copies of other genes, that didn't appear consecutively in the data. 
For three FimA's, we need the derivations:
\begin{equation*}
    N_{7}\Rightarrow N_{1} N_{8}\Rightarrow FimA N_{1} N_{1}\Rightarrow FimA FimA FimA
\end{equation*}
whose product yields a total probability of $0.8$.

For deriving three FimC's we need to first derive $N_{10}\Rightarrow N_{28} N_{29}$ and then:
\begin{equation*}
    N_{29}\Rightarrow N_{3} N_{29}\Rightarrow FimC N_{3} N_{3}\Rightarrow FimC FimC FimC
\end{equation*}
The probability of deriving FimC FimC FimC from $N_{29}$ is $0.16$. Deriving $N_{29}$ from $N_{10}$ instead of $N_{3}$ is also $4$ times less likely, thus the probability for these two duplications is indeed $0.2\cdot0.2=0.04$.
}
\begin{figure} 
    \scalebox{0.25}{\includegraphics{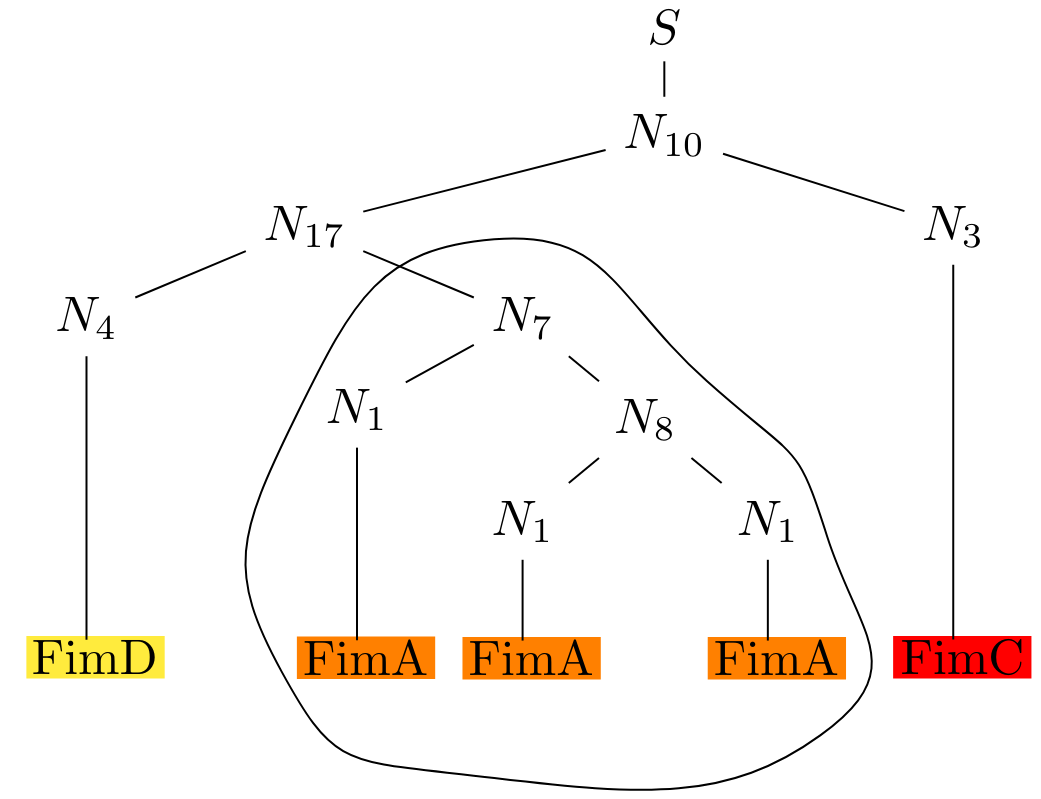}}
    \caption{A parse tree from the grammar learned from the FimACD gene cluster  dataset. The marked sub-tree can grow indefinitely (see Figure~\ref{fig:fimA_dup}), using the production $N7\rightarrow N1 N7$ with a probability of $0.2$.}
    \vspace{-4mm}
    \label{fig:fimACD_tree}
\end{figure}

\begin{figure}
    \scalebox{0.4}{\includegraphics{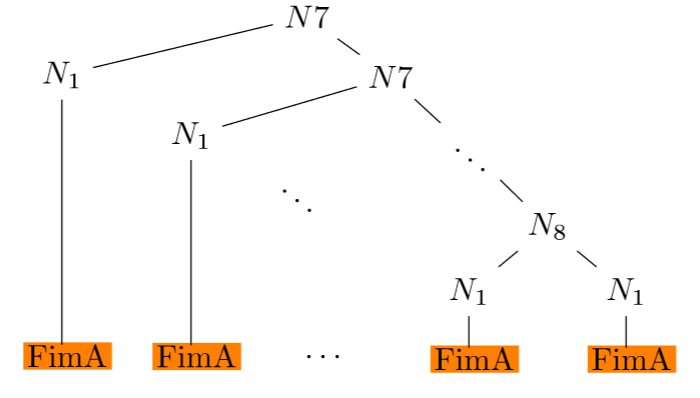}}
    \caption{A right chain formed from $N7$, generating $n$ FimA's with a probability of $0.8\cdot (0.2)^{n-3}$ for $n\geq 3$}
    \vspace{-4mm}
    \label{fig:fimA_dup}
\end{figure}
\commentout{
\begin{figure}
    \scalebox{0.4}{\includegraphics{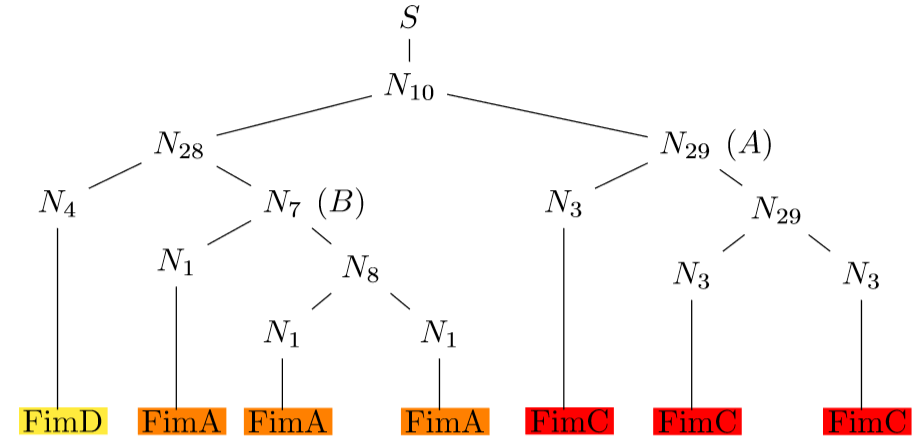}}
    \caption{The tree from figure \ref{fig:fimACD_tree}, with two duplications of the COG FimC}
    \vspace{-4mm}
    \label{fig:fimC_dup}
\end{figure}
}
\subsection{FimACD Grammar}\label{sec:fimacd_grammar}
\begin{tabular}{|c | c|}
Production & Probability\\
\hline
$S \rightarrow  N2 $&0.103\\
\hline
$S \rightarrow  N9 $&0.335\\
\hline
$S \rightarrow  N10 $&0.050\\
\hline
$S \rightarrow  N11 $&0.062\\
\hline
$S \rightarrow  N12 $&0.028\\
\hline
$S \rightarrow  N18 $&0.036\\
\hline
$S \rightarrow  N19 $&0.056\\
\hline
$S \rightarrow  N20 $&0.032\\
\hline
$S \rightarrow  N21 $&0.030\\
\hline
$S \rightarrow  N22 $&0.053\\
\hline
$S \rightarrow  N23 $&0.039\\
\hline
$S \rightarrow  N24 $&0.026\\
\hline
$S \rightarrow  N25 $&0.037\\
\hline
$S \rightarrow  N26 $&0.008\\
\hline
$S \rightarrow  N27 $&0.037\\
\hline
$S \rightarrow  N28 $&0.038\\
\hline
$S \rightarrow  N29 $&0.030\\
\hline
$N1 \rightarrow  \text{FimA} $&1.000\\
\hline
$N2 \rightarrow  N3\,N4 $&0.213\\
\hline
$N2 \rightarrow  N3\, N17 $&0.587\\
\hline
$N2 \rightarrow  N13\, N4 $&0.053\\
\hline
$N2 \rightarrow  N13\, N17 $&0.147\\
\hline
$N3 \rightarrow  \text{FimC} $&1.000\\
\hline
$N4 \rightarrow  \text{FimD} $&1.000\\
\hline
$N5 \rightarrow  \text{CitB} $&1.000\\
\hline
$N6 \rightarrow  N1\,N3 $&0.014\\
\hline
$N6 \rightarrow  N1\, N13 $&0.004\\
\hline
$N6 \rightarrow  N7\, N3 $&0.446\\
\hline
$N6 \rightarrow  N7\, N13 $&0.111\\
\hline
$N6 \rightarrow  N8\, N3 $&0.071\\
\hline
$N6 \rightarrow  N8\, N13 $&0.018\\
\hline
$N6 \rightarrow  N14\, N3 $&0.139\\
\hline
$N6 \rightarrow  N14\, N13 $&0.035\\
\hline
$N6 \rightarrow  N15\, N3 $&0.018\\
\hline
$N6 \rightarrow  N15\, N13 $&0.004\\
\hline
$N6 \rightarrow  N16\, N3 $&0.111\\
\hline
$N6 \rightarrow  N16\, N13 $&0.028\\
\hline
$N7 \rightarrow  N1\, N7 $&0.200\\
\hline
$N7 \rightarrow  N1\, N8 $&0.800\\
\hline
$N8 \rightarrow  N1\, N1 $&1.000\\
\hline
$N9 \rightarrow  N1\, N10 $&0.115\\
\hline
$N9 \rightarrow  N1\, N26 $&0.023\\
\hline
$N9 \rightarrow  N1\, N27 $&0.088\\
\hline
$N9 \rightarrow  N1\, N28 $&0.100\\
\hline
$N9 \rightarrow  N1\, N29 $&0.075\\
\hline
$N9 \rightarrow  N2\, N6 $&0.092\\
\hline
$N9 \rightarrow  N7\, N10 $&0.008\\
\hline
$N9 \rightarrow  N7\, N26 $&0.001\\
\hline
$N9 \rightarrow  N7\, N27 $&0.004\\
\hline
$N9 \rightarrow  N7\, N28 $&0.005\\
\hline
$N9 \rightarrow  N7\, N29 $&0.004\\
\hline
$N9 \rightarrow  N8\, N10 $&0.030\\
\hline
$N9 \rightarrow  N8\, N26 $&0.005\\
\hline
$N9 \rightarrow  N8\, N27 $&0.018\\
\hline
$N9 \rightarrow  N8\, N28 $&0.020\\
\hline
\end{tabular}
\newpage
\begin{tabular}{|c | c|}
$N9 \rightarrow  N8\, N29 $&0.015\\
\hline
$N9 \rightarrow  N11\, N6 $&0.085\\
\hline
$N9 \rightarrow  N12\, N6 $&0.021\\
\hline
$N9 \rightarrow  N15\, N2 $&0.016\\
\hline
$N9 \rightarrow  N15\, N10 $&0.008\\
\hline
$N9 \rightarrow  N15\, N19 $&0.016\\
\hline
$N9 \rightarrow  N15\, N20 $&0.004\\
\hline
$N9 \rightarrow  N15\, N25 $&0.018\\
\hline
$N9 \rightarrow  N16\, N2 $&0.004\\
\hline
$N9 \rightarrow  N16\, N10 $&0.002\\
\hline
$N9 \rightarrow  N16\, N19 $&0.004\\
\hline
$N9 \rightarrow  N16\, N20 $&0.001\\
\hline
$N9 \rightarrow  N16\, N25 $&0.004\\
\hline
$N9 \rightarrow  N18\, N5 $&0.053\\
\hline
$N9 \rightarrow  N18\, N30 $&0.013\\
\hline
$N9 \rightarrow  N19\, N5 $&0.009\\
\hline
$N9 \rightarrow  N19\, N30 $&0.002\\
\hline
$N9 \rightarrow  N20\, N5 $&0.014\\
\hline
$N9 \rightarrow  N20\, N30 $&0.003\\
\hline
$N9 \rightarrow  N21\, N5 $&0.013\\
\hline
$N9 \rightarrow  N21\, N30 $&0.003\\
\hline
$N9 \rightarrow  N22\, N5 $&0.009\\
\hline
$N9 \rightarrow  N22\, N30 $&0.002\\
\hline
$N9 \rightarrow  N23\, N5 $&0.017\\
\hline
$N9 \rightarrow  N23\, N30 $&0.004\\
\hline
$N9 \rightarrow  N24\, N5 $&0.002\\
\hline
$N9 \rightarrow  N24\, N30 $&0.001\\
\hline
$N9 \rightarrow  N25\, N5 $&0.055\\
\hline
$N9 \rightarrow  N25\, N30 $&0.014\\
\hline
$N10 \rightarrow  N4\, N3 $&0.213\\
\hline
$N10 \rightarrow  N4\, N13 $&0.053\\
\hline
$N10 \rightarrow  N17\, N3 $&0.587\\
\hline
$N10 \rightarrow  N17\, N13 $&0.147\\
\hline
$N11 \rightarrow  N1\, N2 $&1.000\\
\hline
$N12 \rightarrow  N7\, N2 $&0.200\\
\hline
$N12 \rightarrow  N8\, N2 $&0.800\\
\hline
$N13 \rightarrow  N3\, N3 $&0.800\\
\hline
$N13 \rightarrow  N3\, N13 $&0.200\\
\hline
$N14 \rightarrow  N1\, N15 $&0.640\\
\hline
$N14 \rightarrow  N1\, N16 $&0.160\\
\hline
$N14 \rightarrow  N7\, N15 $&0.032\\
\hline
$N14 \rightarrow  N7\, N16 $&0.008\\
\hline
$N14 \rightarrow  N8\, N15 $&0.128\\
\hline
$N14 \rightarrow  N8\, N16 $&0.032\\
\hline
$N15 \rightarrow  N7\, N1 $&0.200\\
\hline
$N15 \rightarrow  N8\, N1 $&0.800\\
\hline
$N16 \rightarrow  N7\, N7 $&0.040\\
\hline
$N16 \rightarrow  N7\, N8 $&0.160\\
\hline
$N16 \rightarrow  N8\, N7 $&0.160\\
\hline
$N16 \rightarrow  N8\, N8 $&0.640\\
\hline
$N17 \rightarrow  N4\, N1 $&0.003\\
\hline
$N17 \rightarrow  N4\, N4 $&0.073\\
\hline
$N17 \rightarrow  N4\, N7 $&0.108\\
\hline
$N17 \rightarrow  N4\, N8 $&0.017\\
\hline
$N17 \rightarrow  N4\, N14 $&0.034\\
\hline
\end{tabular}
\newpage
\begin{tabular}{|c | c|}
$N17 \rightarrow  N4\, N15 $&0.004\\
\hline
$N17 \rightarrow  N4\, N16 $&0.027\\
\hline
$N17 \rightarrow  N4\, N17 $&0.200\\
\hline
$N17 \rightarrow  N17\, N1 $&0.010\\
\hline
$N17 \rightarrow  N17\, N7 $&0.297\\
\hline
$N17 \rightarrow  N17\, N8 $&0.048\\
\hline
$N17 \rightarrow  N17\, N14 $&0.093\\
\hline
$N17 \rightarrow  N17\, N15 $&0.012\\
\hline
$N17 \rightarrow  N17\, N16 $&0.074\\
\hline
$N18 \rightarrow  N1\, N25 $&1.000\\
\hline
$N19 \rightarrow  N2\, N1 $&1.000\\
\hline
$N20 \rightarrow  N2\, N7 $&0.800\\
\hline
$N20 \rightarrow  N2\, N16 $&0.200\\
\hline
$N21 \rightarrow  N7\, N25 $&0.200\\
\hline
$N21 \rightarrow  N8\, N25 $&0.800\\
\hline
$N22 \rightarrow  N1\, N19 $&1.000\\
\hline
$N23 \rightarrow  N1\, N20 $&0.800\\
\hline
$N23 \rightarrow  N7\, N20 $&0.040\\
\hline
$N23 \rightarrow  N8\, N20 $&0.160\\
\hline
$N24 \rightarrow  N7\, N19 $&0.200\\
\hline
$N24 \rightarrow  N8\, N19 $&0.800\\
\hline
$N25 \rightarrow  N2\, N8 $&0.800\\
\hline
$N25 \rightarrow  N2\, N15 $&0.200\\
\hline
$N26 \rightarrow  N2\, N14 $&1.000\\
\hline
$N27 \rightarrow  N10\, N1 $&1.000\\
\hline
$N28 \rightarrow  N10\, N7 $&0.640\\
\hline
$N28 \rightarrow  N10\, N14 $&0.200\\
\hline
$N28 \rightarrow  N10\, N16 $&0.160\\
\hline
$N29 \rightarrow  N10\, N8 $&0.800\\
\hline
$N29 \rightarrow  N10\, N15 $&0.200\\
\hline
$N30 \rightarrow  N5\, N5 $&0.800\\
\hline
$N30 \rightarrow  N5\, N30 $&0.200\\
\hline
\end{tabular}
\newpage

\stepcounter{section}
\clearpage
\bibliography{bib}

\begin{thebibliography}{44}
\providecommand{\natexlab}[1]{#1}
\providecommand{\url}[1]{\texttt{#1}}
\providecommand{\urlprefix}{URL }
\expandafter\ifx\csname urlstyle\endcsname\relax
  \providecommand{\doi}[1]{doi:\discretionary{}{}{}#1}\else
  \providecommand{\doi}{doi:\discretionary{}{}{}\begingroup
  \urlstyle{rm}\Url}\fi

\bibitem[{Abe and Warmuth(1992)}]{AbeW92}
Abe, N.; and Warmuth, M.~K. 1992.
\newblock On the Computational Complexity of Approximating Distributions by
  Probabilistic Automata.
\newblock \emph{Machine Learning} 9: 205--260.

\bibitem[{Abney, McAllester, and Pereira(1999)}]{abney1999relating}
Abney, S.; McAllester, D.; and Pereira, F. 1999.
\newblock Relating probabilistic grammars and automata.
\newblock In \emph{Proceedings of the 37th Annual Meeting of the Association
  for Computational Linguistics}, 542--549.

\bibitem[{Alvarez-Ortega, Olivares, and Mart{\'\i}nez(2013)}]{alvarez2013rnd}
Alvarez-Ortega, C.; Olivares, J.; and Mart{\'\i}nez, J.~L. 2013.
\newblock RND multidrug efflux pumps: what are they good for?
\newblock \emph{Frontiers in microbiology} 4: 7.

\bibitem[{Angluin(1987)}]{Angluin87}
Angluin, D. 1987.
\newblock Learning Regular Sets from Queries and Counterexamples.
\newblock \emph{Inf. Comput.} 75(2): 87--106.

\bibitem[{Angluin(1990)}]{Angluin90}
Angluin, D. 1990.
\newblock Negative Results for Equivalence Queries.
\newblock \emph{Machine Learning} 5: 121--150.

\bibitem[{Angluin and Kharitonov(1995)}]{AngluinK95}
Angluin, D.; and Kharitonov, M. 1995.
\newblock When Won't Membership Queries Help?
\newblock \emph{J. Comput. Syst. Sci.} 50(2): 336--355.

\bibitem[{Baker(1979)}]{Baker79}
Baker, J.~K. 1979.
\newblock Trainable grammars for speech recognition.
\newblock In Klatt, D.~H.; and Wolf, J.~J., eds., \emph{Speech Communication
  Papers for the 97th Meeting of the Acoustical Society of America}, 547--550.

\bibitem[{Bergadano and Varricchio(1996)}]{BergadanoV96}
Bergadano, F.; and Varricchio, S. 1996.
\newblock Learning Behaviors of Automata from Multiplicity and Equivalence
  Queries.
\newblock \emph{{SIAM} J. Comput.} 25(6): 1268--1280.

\bibitem[{Bergeron(2008)}]{bergeron2008formal}
Bergeron, A. 2008.
\newblock Formal models of gene clusters .

\bibitem[{Booth and Lueker(1976)}]{booth1976testing}
Booth, K.~S.; and Lueker, G.~S. 1976.
\newblock Testing for the consecutive ones property, interval graphs, and graph
  planarity using PQ-tree algorithms.
\newblock \emph{Journal of Computer and System Sciences} 13(3): 335--379.

\bibitem[{Chomsky(1956)}]{Chomsky56}
Chomsky, N. 1956.
\newblock Three models for the description of language.
\newblock \emph{{IRE} Trans. Inf. Theory} 2(3): 113--124.
\newblock \doi{10.1109/TIT.1956.1056813}.
\newblock \urlprefix\url{https://doi.org/10.1109/TIT.1956.1056813}.

\bibitem[{Church(1988)}]{church-1988-stochastic}
Church, K.~W. 1988.
\newblock A Stochastic Parts Program and Noun Phrase Parser for Unrestricted
  Text.
\newblock In \emph{Second Conference on Applied Natural Language Processing},
  136--143. Austin, Texas, USA: Association for Computational Linguistics.
\newblock \doi{10.3115/974235.974260}.
\newblock \urlprefix\url{https://www.aclweb.org/anthology/A88-1019}.

\bibitem[{Cohen and Rothblum(1993)}]{cohen1993nonnegative}
Cohen, J.~E.; and Rothblum, U.~G. 1993.
\newblock Nonnegative ranks, decompositions, and factorizations of nonnegative
  matrices.
\newblock \emph{Linear Algebra and its Applications} 190: 149--168.

\bibitem[{de~la Higuera(2010)}]{delaHiguera}
de~la Higuera, C. 2010.
\newblock \emph{Grammatical Inference: Learning Automata and Grammars}.
\newblock USA: Cambridge University Press.
\newblock ISBN 0521763169.

\bibitem[{Drewes and H{\"o}gberg(2007)}]{drewes2007query}
Drewes, F.; and H{\"o}gberg, J. 2007.
\newblock Query learning of regular tree languages: How to avoid dead states.
\newblock \emph{Theory of Computing Systems} 40(2): 163--185.

\bibitem[{Fani, Brilli, and Lio(2005)}]{fani2005origin}
Fani, R.; Brilli, M.; and Lio, P. 2005.
\newblock The origin and evolution of operons: the piecewise building of the
  proteobacterial histidine operon.
\newblock \emph{Journal of molecular evolution} 60(3): 378--390.

\bibitem[{Fondi, Emiliani, and Fani(2009)}]{fondi2009origin}
Fondi, M.; Emiliani, G.; and Fani, R. 2009.
\newblock Origin and evolution of operons and metabolic pathways.
\newblock \emph{Research in microbiology} 160(7): 502--512.

\bibitem[{Ge, Yamada, and Zgurskaya(2009)}]{ge2009c}
Ge, Q.; Yamada, Y.; and Zgurskaya, H. 2009.
\newblock The C-terminal domain of AcrA is essential for the assembly and
  function of the multidrug efflux pump AcrAB-TolC.
\newblock \emph{Journal of bacteriology} 191(13): 4365--4371.

\bibitem[{Gold(1978)}]{Gold78}
Gold, E.~M. 1978.
\newblock Complexity of Automaton Identification from Given Data.
\newblock \emph{Information and Control} 37(3): 302--320.

\bibitem[{Grate(1995)}]{Grate95}
Grate, L. 1995.
\newblock Automatic {RNA} Secondary Structure Determination with Stochastic
  Context-Free Grammars.
\newblock In Rawlings, C.~J.; Clark, D.~A.; Altman, R.~B.; Hunter, L.;
  Lengauer, T.; and Wodak, S.~J., eds., \emph{Proceedings of the Third
  International Conference on Intelligent Systems for Molecular Biology,
  Cambridge, United Kingdom, July 16-19, 1995}, 136--144. {AAAI}.
\newblock \urlprefix\url{http://www.aaai.org/Library/ISMB/1995/ismb95-017.php}.

\bibitem[{Habrard and Oncina(2006)}]{habrard2006learning}
Habrard, A.; and Oncina, J. 2006.
\newblock Learning multiplicity tree automata.
\newblock In \emph{International Colloquium on Grammatical Inference},
  268--280. Springer.

\bibitem[{Hayashi et~al.(2016)Hayashi, Nakashima, Sakurai, Kitagawa, Yamasaki,
  Nishino, and Yamaguchi}]{hayashi2016acrb}
Hayashi, K.; Nakashima, R.; Sakurai, K.; Kitagawa, K.; Yamasaki, S.; Nishino,
  K.; and Yamaguchi, A. 2016.
\newblock AcrB-AcrA fusion proteins that act as multidrug efflux transporters.
\newblock \emph{Journal of bacteriology} 198(2): 332--342.

\bibitem[{Hopcroft and Ullman(1979)}]{HopcroftUllman79}
Hopcroft, J.~E.; and Ullman, J.~D. 1979.
\newblock \emph{Introduction to Automata Theory, Languages, and Computation}.
\newblock Addison-Wesley Publishing Company.

\bibitem[{Kasami(1966)}]{kasami1966efficient}
Kasami, T. 1966.
\newblock An efficient recognition and syntax-analysis algorithm for
  context-free languages.
\newblock \emph{Coordinated Science Laboratory Report no. R-257} .

\bibitem[{Kobylka et~al.(2020)Kobylka, Kuth, M{\"u}ller, Geertsma, and
  Pos}]{kobylka2020acrb}
Kobylka, J.; Kuth, M.~S.; M{\"u}ller, R.~T.; Geertsma, E.~R.; and Pos, K.~M.
  2020.
\newblock AcrB: a mean, keen, drug efflux machine.
\newblock \emph{Annals of the New York Academy of Sciences} 1459(1): 38--68.

\bibitem[{Labedan and Riley(1995)}]{labedan1995widespread}
Labedan, B.; and Riley, M. 1995.
\newblock Widespread protein sequence similarities: origins of Escherichia coli
  genes.
\newblock \emph{Journal of bacteriology} 177(6): 1585--1588.

\bibitem[{Lalanne et~al.(2018)Lalanne, Taggart, Guo, Herzel, Schieler, and
  Li}]{lalanne2018evolutionary}
Lalanne, J.-B.; Taggart, J.~C.; Guo, M.~S.; Herzel, L.; Schieler, A.; and Li,
  G.-W. 2018.
\newblock Evolutionary convergence of pathway-specific enzyme expression
  stoichiometry.
\newblock \emph{Cell} 173(3): 749--761.

\bibitem[{Landau, Parida, and Weimann(2005)}]{landau2005gene}
Landau, G.~M.; Parida, L.; and Weimann, O. 2005.
\newblock Gene proximity analysis across whole genomes via pq trees1.
\newblock \emph{Journal of Computational Biology} 12(10): 1289--1306.

\bibitem[{Lari and Young(1990)}]{LariY90}
Lari, K.; and Young, S.~J. 1990.
\newblock The estimation of stochastic context-free grammars using the
  Inside-Outside algorithm.
\newblock \emph{Computer Speech and Language} 4: 35--56.

\bibitem[{Levy and Joshi(1978)}]{LevyJ78}
Levy, L.~S.; and Joshi, A.~K. 1978.
\newblock Skeletal structural descriptions.
\newblock \emph{Information and Control} 39(2): 192 -- 211.
\newblock ISSN 0019-9958.
\newblock \doi{https://doi.org/10.1016/S0019-9958(78)90849-5}.
\newblock
  \urlprefix\url{http://www.sciencedirect.com/science/article/pii/S0019995878908495}.

\bibitem[{Lewis(1951)}]{lewis1951pseudoallelism}
Lewis, E.~B. 1951.
\newblock Pseudoallelism and gene evolution.
\newblock In \emph{Cold Spring Harbor Symposia on Quantitative Biology},
  volume~16, 159--174. Cold Spring Harbor Laboratory Press.

\bibitem[{Rabiner(1989)}]{Rabiner89}
Rabiner, L.~R. 1989.
\newblock A tutorial on hidden Markov models and selected applications in
  speech recognition.
\newblock In \emph{PROCEEDINGS OF THE IEEE}, 257--286.

\bibitem[{Regis(2016)}]{Regis2016}
Regis, R.~G. 2016.
\newblock On the properties of positive spanning sets and positive bases.
\newblock \emph{Optimization and Engineering} 17(1): 229--262.
\newblock ISSN 1573-2924.
\newblock \doi{10.1007/s11081-015-9286-x}.
\newblock \urlprefix\url{https://doi.org/10.1007/s11081-015-9286-x}.

\bibitem[{Sakakibara(1988)}]{Sakakibara88}
Sakakibara, Y. 1988.
\newblock Learning Context-Free Grammars from Structural Data in Polynomial
  Time.
\newblock In \emph{Proceedings of the First Annual Workshop on Computational
  Learning Theory, {COLT} '88, Cambridge, MA, USA, August 3-5, 1988}, 330--344.
\newblock \urlprefix\url{http://dl.acm.org/citation.cfm?id=93109}.

\bibitem[{Sakakibara(1992)}]{Sakakibara92}
Sakakibara, Y. 1992.
\newblock Efficient learning of context-free grammars from positive structural
  examples.
\newblock \emph{Inform. Comput.} 97: 23--60.

\bibitem[{Shi et~al.(2019)Shi, Chen, Yu, Bell, Wang, Forrester, Villarreal,
  Jakana, Du, Luisi et~al.}]{shi2019situ}
Shi, X.; Chen, M.; Yu, Z.; Bell, J.~M.; Wang, H.; Forrester, I.; Villarreal,
  H.; Jakana, J.; Du, D.; Luisi, B.~F.; et~al. 2019.
\newblock In situ structure and assembly of the multidrug efflux pump
  AcrAB-TolC.
\newblock \emph{Nature communications} 10(1): 1--6.

\bibitem[{Smith and Johnson(2007)}]{smith2007weighted}
Smith, N.~A.; and Johnson, M. 2007.
\newblock Weighted and probabilistic context-free grammars are equally
  expressive.
\newblock \emph{Computational Linguistics} 33(4): 477--491.

\bibitem[{Svetlitsky et~al.(2019)Svetlitsky, Dagan, Chalifa-Caspi, and
  Ziv-Ukelson}]{svetlitsky2019csbfinder}
Svetlitsky, D.; Dagan, T.; Chalifa-Caspi, V.; and Ziv-Ukelson, M. 2019.
\newblock CSBFinder: discovery of colinear syntenic blocks across thousands of
  prokaryotic genomes.
\newblock \emph{Bioinformatics} 35(10): 1634--1643.

\bibitem[{Svetlitsky, Dagan, and Ziv-Ukelson(2020)}]{csbfinder-s}
Svetlitsky, D.; Dagan, T.; and Ziv-Ukelson, M. 2020.
\newblock Discovery of multi-operon colinear syntenic blocks in microbial
  genomes.
\newblock \doi{btaa503}.
\newblock In press.

\bibitem[{Tatusov et~al.(2000)Tatusov, Galperin, Natale, and
  Koonin}]{tatusov2000cog}
Tatusov, R.~L.; Galperin, M.~Y.; Natale, D.~A.; and Koonin, E.~V. 2000.
\newblock The COG database: a tool for genome-scale analysis of protein
  functions and evolution.
\newblock \emph{Nucleic acids research} 28(1): 33--36.

\bibitem[{Weiss, Goldberg, and Yahav(2019)}]{WeissGY19}
Weiss, G.; Goldberg, Y.; and Yahav, E. 2019.
\newblock Learning Deterministic Weighted Automata with Queries and
  Counterexamples.
\newblock In Wallach, H.~M.; Larochelle, H.; Beygelzimer, A.;
  d'Alch{\'{e}}{-}Buc, F.; Fox, E.~B.; and Garnett, R., eds., \emph{Advances in
  Neural Information Processing Systems 32: Annual Conference on Neural
  Information Processing Systems 2019, NeurIPS 2019, 8-14 December 2019,
  Vancouver, BC, Canada}, 8558--8569.
\newblock
  \urlprefix\url{http://papers.nips.cc/paper/9062-learning-deterministic-weighted-automata-with-queries-and-counterexamples}.

\bibitem[{Weston et~al.(2018)Weston, Sharma, Ricci, and
  Piddock}]{weston2018regulation}
Weston, N.; Sharma, P.; Ricci, V.; and Piddock, L.~J. 2018.
\newblock Regulation of the AcrAB-TolC efflux pump in Enterobacteriaceae.
\newblock \emph{Research in microbiology} 169(7-8): 425--431.

\bibitem[{Winter et~al.(2016)Winter, Jahn, Wehner, Kuchenbecker, Marz, Stoye,
  and B{\"o}cker}]{winter2016finding}
Winter, S.; Jahn, K.; Wehner, S.; Kuchenbecker, L.; Marz, M.; Stoye, J.; and
  B{\"o}cker, S. 2016.
\newblock Finding approximate gene clusters with Gecko 3.
\newblock \emph{Nucleic Acids Research} 44(20): 9600--9610.

\bibitem[{Younger(1967)}]{younger1967recognition}
Younger, D.~H. 1967.
\newblock Recognition and parsing of context-free languages in time n3.
\newblock \emph{Information and control} 10(2): 189--208.

\end{thebibliography}
}
\end{document}